\documentclass[11pt]{article}
\usepackage{mathrsfs}
\usepackage{natbib,graphicx,setspace,lscape,longtable}
\usepackage{mathrsfs,amsmath,amsthm,amssymb,color}
\usepackage{natbib,epsfig,graphicx,pdfpages}
\usepackage{rotating}
\usepackage{subfigure}
\usepackage{enumerate}
\usepackage{graphicx}
\usepackage{algorithmic}
\usepackage{algorithm}

\usepackage{url}
\usepackage{hyperref}
\usepackage[table]{xcolor}
\usepackage{colortbl}
\bibpunct{(}{)}{;}{a}{,}{,}

\newtheorem{theorem}{Theorem}
\newtheorem{lemma}{Lemma}
\newtheorem{proposition}{Proposition}
\newtheorem{remark}{Remark}
\usepackage{amsthm}

\newtheorem{assumption}{{Assumption}}
\def\beq{\begin{equation}}
\def\eeq{\end{equation}}
\def\beqr{\begin{eqnarray}}
\def\eeqr{\end{eqnarray}}
\def\beqrs{\begin{eqnarray*}}
\def\eeqrs{\end{eqnarray*}}
\def\bet{\begin{theorem}}
\def\eet{\end{theorem}}
\def\bel{\begin{lemma}}
\def\eel{\end{lemma}}
\def\bep{\begin{proposition}}
\def\eep{\end{proposition}}
\def\bg{\begin{figure}[tbph]\begin{center}}
\def\eg{\end{center}\end{figure}}

\def\bc{\begin{center}}
\def\ec{\end{center}}

\def\wh{\widehat}

\def\cov{\mbox{Cov}}

\def\diag{\mbox{diag}}

\numberwithin{equation}{section}

\newcommand{\Cov}{\textnormal{Cov}}

\newcommand{\bA}{{\mathbf A}}
\newcommand{\bB}{{\mathbf B}}

\newcommand{\bE}{{\mathbf E}}
\newcommand{\bG}{{\mathbf G}}
\newcommand{\bH}{{\mathbf H}}
\newcommand{\bI}{{\mathbf I}}

\newcommand{\bL}{{\mathbf L}}
\newcommand{\bM}{{\mathbf M}}

\newcommand{\bQ}{{\mathbf Q}}
\newcommand{\bP}{{\mathbf P}}
\newcommand{\bR}{{\mathbf R}}
\newcommand{\bS}{{\mathbf S}}

\newcommand{\bU}{{\mathbf U}}
\newcommand{\bV}{{\mathbf V}}

\newcommand{\bZ}{{\mathbf Z}}
\newcommand{\ba}{{\mathbf a}}
\newcommand{\bb}{{\mathbf b}}

 \newcommand{\bfc}{{\mathbf c}}
\newcommand{\be}{{\mathbf e}}
\newcommand{\bff}{{\mathbf f}}
 
\newcommand{\bh}{{\mathbf h}}

\newcommand{\br}{{\mathbf r}}

\newcommand{\bu}{{\mathbf u}}

\newcommand{\bw}{{\mathbf w}}
\newcommand{\bx}{{\mathbf x}}
\newcommand{\by}{{\mathbf y}}
\newcommand{\bz}{{\mathbf z}}

\newcommand{\bfeta}  {\boldsymbol{\eta}}

\newcommand{\bSigma}{\boldsymbol{\Sigma}}
\newcommand{\bDelta}{\boldsymbol{\Delta}}

\newcommand{\bve}{\mbox{\boldmath$\varepsilon$}}

\newcommand{\bTheta} {\boldsymbol{\Theta}}
\newcommand{\bPhi} {\boldsymbol{\Phi}}

\newcommand{\bxi} {\boldsymbol{\xi}}

\newcommand{\bGamma} {\boldsymbol{\Gamma}}

\newcommand{\bD}{{\mathbf D}}

\newcommand{\ve}{{\varepsilon}}
\renewcommand{\epsilon}{{\ve}}
\renewcommand{\hat}{\widehat}

\def\JRSSB{{\sl Journal of the Royal Statistical Society}, {\bf B}}

\def\BKA{{\sl Biometrika}}
\def\JASA{{\sl Journal of the American Statistical Association}}

\begin{document}

\title{Modeling High-Dimensional Dependent Data in the Presence of 
Many Explanatory Variables and Weak Signals}
\author{
Zhaoxing Gao$^1$ and Ruey S. Tsay$^2$\footnote{Corresponding author: \href{mailto:ruey.tsay@chicagobooth.edu}{ruey.tsay@chicagobooth.edu} (R.S. Tsay).  Booth School of Business, University of Chicago, 5807 South Woodlawn Avenue, Chicago, IL, 60637, USA.}  \\
$^1$School of Mathematical Sciences, University of Electronic Science and Technology of China\\
$^2$Booth School of Business, University of Chicago
}

 \date{}

\maketitle

\begin{abstract}
{This article considers a novel and widely applicable approach to  modeling high-dimensional dependent data when a large number of explanatory 
variables are available and the signal to noise ratio is low. 
We postulate that a $p$-dimensional response series is the sum of a linear regression with many observable explanatory variables and an error term driven by some latent common factors and an idiosyncratic noise. The common factors have dynamic dependence whereas the covariance matrix of the idiosyncratic noise can have diverging eigenvalues to handle the 
situation of low 
signal to noise ratio commonly encountered in applications.
The regression coefficient matrix is estimated using penalized methods when the dimension of the predictors is high. We apply factor modeling to the regression residuals, employ a high-dimensional white noise testing procedure to determine the number of 
common factors, and adopt a projected Principal Component Analysis when the 
signal to noise ratio is low. 
We establish asymptotic properties of the proposed method, both for fixed and diverging numbers of regressors, as 
$p$ and the sample size 
$T$ approach infinity. Finally, we use simulations and empirical applications to demonstrate the efficacy of the proposed approach in finite samples. }

\end{abstract}


\newpage

\section{Introduction}

{The availability of large-scale and serially dependent datasets presents both opportunities and challenges for modern data analysts. Modeling and forecasting high-dimensional time series have emerged across various scientific fields, including economics, finance, social and environmental studies, and Internet data processing. In many cases, the number of variables can be as large as, or even exceed, the number of observations, making statistical inference particularly challenging. Addressing such issues has 
attracted much recent research interest and has become a central focus in contemporary data analysis. In many applications, a large number of explanatory variables are also available. For instance, in studying daily COVID-19 case counts across the 50 States of the U.S., one might also incorporate variables that are believed to influence the spreading of the disease, such as daily temperature, population size, and vaccination rates. In portfolio optimization and risk management, the number of assets under study often reaches hundreds or thousands, and there also exist various financial factors, such as market conditions, interest rates, and oil price shocks, that can affect asset returns. Similarly, in climate modeling, researchers often need to analyze high-dimensional time series data including temperature, precipitation, wind speed, and other atmospheric variables at multiple geographical locations, where local features such as geographic elevation or oceanic currents may play significant roles in predicting long-term climate patterns.}

{Arguably the most commonly used model to analyze time series 
data with explanatory variables is the regression model with time series errors. 
However, for modern large-scale high-dimensional time series 
with many explanatory variables, the conventional model becomes 
inadequate, because 
dimension reduction or structural simplification becomes necessary. 
A huge amount of literature is available in the dimension reduction of dependent data. 
Indeed, numerous methods have been developed for the analysis of multivariate or high-dimensional dependent data without exogenous variables. See, for example, \cite{Hanetal2020} and \cite{gaotsay2018a,gaotsay2020a,gaotsay2018b,gaotsay2021c,gaotsay2021}, among many others.  For factor analysis of dependent data 
without exogenous predictors, readers are referred to \cite{StockWatson_2005}, \cite{BaiNg_Econometrica_2002}, \cite{forni2000,forni2005}, \cite{LamYaoBathia_Biometrika_2011}, \cite{lamyao2012}, \cite{changguoyao2015}, and \cite{chentsaychen2018}. 
But the complexity of dynamic dependencies in high-dimensional dependent data with a large number of explanatory variables deserves further exploration. In particular, jointly considering the effect of explanatory variables and 
extracting dynamic information from such data plays a vital role in effectively modeling and forecasting serially dependent systems.}

{The goal of this paper is to study linear regression models for 
high-dimensional time series 
with many explanatory. We consider high-dimensional 
linear regression models with many predictors using regularized estimation 
and factor models. In addition, we allow the covariance matrix of the 
idiosyncratic noise to have diverging eigenvalues, which give rise to 
a low signal-to-noise ratio commonly seen in applications. 
Our study puts together high-dimensional regression and 
factor modeling under a unified framework. 
The work marks an important extension} of \cite{gaotsay2018b}, which introduces a factor model by treating a 
$p$-dimensional vector time series as a nonsingular linear transformation of some common factors and idiosyncratic components. In their model, the factor process is dynamically dependent whereas the idiosyncratic component is a white noise process. In particular, the largest eigenvalues of the covariance matrix of the idiosyncratic component can diverge to infinity as the dimension 
$p$ increases, reflecting the case of low signal-to-noise ratio commonly observed in financial and economic data. 
{More specifically, in contrast to \cite{gaotsay2018b}, 
which focuses on the original 
time series data, our approach assumes that a large number of 
explanatory variables are available and the error term in the 
high-dimensional regression assumes a factor model structure 
with prominent noise effects.} This framework not only acknowledges the fundamental assumption in linear regression that the error term should behave 
as a white noise sequence but also aims to extract potential dynamic information from the residuals of high-dimensional regression, as they may still capture significant dynamic dependence in the data despite the high dimensionality.

The concept of integrating observable and latent factors has been considered in \cite{changguoyao2015}. {However, our approach distinguishes itself by considering both a fixed and diverging number of observable regressors and establishing the theoretical underpinnings for the setting. Furthermore, our factor process accommodates prominent noise effects, and we propose a prediction procedure that leverages both the regressors and the dynamic factors extracted from the data.}

{The integration of high-dimensional linear regression and 
factor models dramatically increase the applicability of individual models. 
The combined model is particularly relevant in modern big data 
environment. 
From a factor modeling perspective, observable regressors can be considered as known factors that drive the dynamics of the data jointly with latent factors inherent in the model. For instance, temperature can serve as a significant factor in forecasting household electricity consumption. In finance, the market index is often treated as a common factor for pricing various assets within the Capital Asset Pricing Model (CAPM) (see \cite{sharpe1964}). Additionally, the regressors can represent a high-dimensional vector that can explain  
the data variability and improve the accuracy in forecasting the original data. 
From linear regression perspective, the use of factor models with diverging 
noise effect relaxes the independence assumption and achieves a high degree of 
dimension reduction. 
}

We establish asymptotic properties of the proposed method for both fixed and diverging numbers of regressors as the dimension 
$p$ and sample size 
$T$ approach infinity. Our findings demonstrate that the factor modeling framework is asymptotically adaptive to unknown regression coefficients, indicating that the convergence rates for estimating the factor loading space and the factor process align with those achievable under the assumption of known regression coefficients.
To evaluate the performance of the proposed method, we conduct analyses using both simulated data and real-world examples. The empirical applications suggest that our method effectively models stock returns, with additional latent factors significantly enhancing the predictive accuracy for asset returns.

The contributions of this paper are multifaceted. First, the proposed method 
{extends} the work of \cite{gaotsay2018b} by incorporating observable common factors, a feature commonly observed in empirical applications such as financial asset pricing, thereby significantly extending the applicability of the model. Second, our approach offers a refined methodology for extracting dynamic information from high-dimensional regression models while mitigating potential noise effects. This dynamic information captures time series dynamics, which can lead to more accurate forecasts. Third, we establish a theoretical framework demonstrating that latent factor models with prominent noise effects can adapt effectively to high-dimensional regression contexts, provided that appropriate regularization techniques are employed for estimating the regression coefficients in the initial step. This aspect represents an 
important theoretical contribution of our study.

The rest of the paper is organized as follows. 
We introduce the proposed model and 
estimation methodology in Section \ref{sec2} and 
study the theoretical properties of the proposed model and its associated 
 estimates in Section \ref{sec3}.  
Numerical studies with both simulated and real data sets are given in Section \ref{sec4}, and 
Section \ref{sec5} provides some concluding remarks. 
All technical proofs are given in an Appendix. Throughout the article,
 we use the following notation. For a $p\times 1$ vector
$\bu=(u_1,..., u_p)',$  $||\bu||_2 =\|\bu'\|_2= (\sum_{i=1}^{p} u_i^2)^{1/2} $
is the Euclidean norm, $\|\bu\|_\infty=\max_{1\leq i\leq p}|u_i|$ is the $\ell_\infty$-norm, and $\bI_p$ denotes a $p\times p$ identity matrix. For a matrix $\bH=(h_{ij})$, $\|\bH\|_1=\max_j\sum_i|h_{ij}|$, $|\bH|_\infty=\max_{i,j}|h_{ij}|$,  $\|\bH
\|_F=\sqrt{\sum_{i,j}h_{ij}^2}$ is the Frobenius norm, $\|\bH
\|_2=\sqrt{\lambda_{\max} (\bH' \bH ) }$ is the operator norm, where
$\lambda_{\max} (\cdot) $ denotes for the largest eigenvalue of a matrix, and $\|\bH\|_{\min}$ is the square root of the minimum non-zero eigenvalue of $\bH\bH'$. The superscript ${'}$ denotes the 
transpose of  a vector or matrix. We also use the notation $a\asymp b$ to denote $a=O(b)$and $b=O(a)$.

\section{Models and Methodology}\label{sec2}
\subsection{Setting}
Let $\by_t=(y_{1,t},...,y_{p,t})'\in R^p$ and  $\bz_t=(z_{1,t},...,z_{m,t})'\in R^m$ be observable  $p$-dimensional and $m$-dimensional vector time series,  respectively.  We consider the following regression model
\begin{equation}\label{reg}
\by_t=\bB\bz_t+\bfeta_t, 
\end{equation}
{where $\bB$ is a $p\times m$ coefficient matrix and 
$\bfeta_t$ is a 
$p$-dimensional stationary process denoting the regression-adjusted underlying 
process of the data $\by_t$.} In theory, we may assume the dimensions 
$p\asymp m$; however, we focus on the case where 
$m/p\rightarrow 0$, as the number of common regressors is often fixed or 
grows at a slower rate than the dimension $p$. 
{In many applications, the predictor vector $\bz_t$ are 
unable to account for all the dynamic dependence in $\by_t$. For instance, 
some relevant predictors may be missing or the relationship may not 
be linear. }
To further extract the dynamic dependence from the data, we assume that 
$\bfeta_t$  follows a factor model in \cite{gaotsay2018b} and can be  written as follows:
\begin{equation}\label{eta}
\bfeta_t=\bL\left[\begin{array}{c}
\bff_t\\
\bve_t
\end{array}\right]=\bL_1\bff_t+\bL_2\bve_t,
\end{equation}
{where $\bL$ is a $p\times p$ nonsingular transformation matrix, 
$\bff_t$ is an $r\times 1$ latent factor process that captures all the dynamical dependence in $\bfeta_t$, and $\bve_t$ is a $v$-dimensional white noise process with $\cov(\bff_t)=\bI_r$, $\cov(\bve_t)=\bI_v$, and $r+v=p$.}  As discussed in \cite{gaotsay2018b},  Model (\ref{eta}) can be rewritten as $\bL^{-1}\by_t=(\bff_t',\bve_t')'$, which can be constructed via Canonical Correlation Analysis (CCA) between $\by_t$ and its past lagged variables 
provided that $p$ is much smaller than the sample size. 
See \cite{anderson2003} for details of CCA. 
Using the terminology of \cite{TiaoTsay_1989}, $\bff_t$ and $\bve_t$ are scalar components of $\bfeta_t$.  Unlike the traditional factor models, which assume $\bff_{t}$ and $\bve_{s}$ are uncorrelated for any $t$ and $s$, we only require $\Cov(\bff_t,\bve_{t+j}) = 0$ for $j \geq 0$ in this article. The asymptotic properties shown below can be simplified if $\bff_t$ and $\bve_t$ are uncorrelated across all time lags, as discussed in Section 3 below.

In Models (\ref{reg})--(\ref{eta}), only 
$\by_t$
  and 
$\bz_t$
  are observable. 
The matrices $\bL_1$  and $\bL_2$
  are unknown loading matrices that capture the strengths of the factors and idiosyncratic terms, respectively. Consistent with traditional factor modeling literature, we assume that the 
$r$ singular values of 
$\bL_1$
  diverge as the system expands. To address the prominent noise effects encountered in many applications discussed in the Introduction, we also assume that the 
$s$ largest singular values of 
$\bL_2$
  diverge, see \cite{gaotsay2018b}. The number of latent factors 
$r$ and the number of diverging noise components 
$s$ are treated as unknown (but fixed) constants. Our objective is to estimate the regression matrix 
$\bB$, the factor loading matrix 
$\bL_1$, the number of factors 
$r$, and to recover the factor process 
$\bff_t$, particularly in the presence of significant noise effects, while allowing the dimensions 
$p$ and/or 
$m$ to diverge as the sample size 
$T$ increases.


Note that $(\bL_1,\bff_t)$ and $(\bL_2,\bve_t)$ are not uniquely defined;  only the linear space spanned by the columns of $\bL_1$ (and $\bL_2$), denoted by $\mathcal{M}(\bL_1)$ (and $\mathcal{M}(\bL_2)$), can be uniquely defined. Therefore, following the procedure in \cite{gaotsay2018b}, we express $\bL_1=\bA_1\bQ_1$ and $\bL_2=\bA_2\bQ_2$, where $\bA_1$ and $\bA_2$ are two semi-orthogonal matrices satisfying $\bA_1'\bA_1=\bI_{r}$ and $\bA_2'\bA_2=\bI_{v}$. The column spaces of these matrices correspond to $\mathcal{M}(\bL_1)$ and $\mathcal{M}(\bL_2)$, respectively. This representation can be achieved through QR decomposition or singular value decomposition. Let $\bx_t=\bQ_1\bff_t$ and $\be_t=\bQ_2\bve_t$, then Models (\ref{reg}) and (\ref{eta}) can be written as
\begin{equation}\label{full}
\by_t=\bB\bz_t+\bfeta_t=\bB\bz_t+\bA_1\bx_t+\bA_2\be_t,
\end{equation}
where $\bA_1$ is a new factor loading matrix and $\bx_t$ is the associated factor process.  As a result, the eigenvalues of $\cov(\bx_t)$ and the top $s$ eigenvalues of $\cov(\bve_t)$ are diverging.  Our goal now is to estimate the regression matrix $\bB$, the factor loading space $\mathcal{M}(\bA_1)$, the number of factors $r$, and to recover the factor process $\bx_t$. In addition, we consider the simple case that $\bz_t$ and $\bfeta_t$ are uncorrelated, which ensures that the coefficient matrix $\bB$ is identifiable.  When $\bz_t$ and $\bx_t$ are correlated implying the existence of endogeneity,  the regression coefficient matrix and the factor loading space can still be identified so long as some proper instrumental variables are employed. See, for example, \cite{changguoyao2015}. We do not consider this issue in this paper.


\subsection{Estimation}
\subsubsection{Estimation With a Known Number of Factors}
We first assume that the number of factors $r$ is known, and will outline a way to identify $r$ in Section~\ref{sec2.2.2} below.  The proposed methodology is as follows. 

When  the number of regressors $m$ is small, the estimation of $\bB$ can be treated as a standard least-squares problem since $\Cov(\bz_t,\bfeta_t)={\bf 0}$.  Write $\bB=(\bb_1,...,\bb_p)'$, the least-squares estimator (LSE) for $\bB$ can be expressed as $\wh\bB=(\wh\bb_1,...,\wh\bb_p)'$,  where
\begin{equation}
\wh\bb_i=\left(\frac{1}{T}\sum_{t=1}^T\bz_t\bz_t'\right)^{-1}\left(\frac{1}{T}\sum_{t=1}^Ty_{i,t}\bz_t\right),
\end{equation}
where $y_{i,t}$ is the $i$-th component of $\by_t$.  When $m$ is large in relation to the sample size $T$, we apply Lasso regression to each $y_{i,t}$ to estimate $\bb_i$ and perform variable selection. Specifically, for each $i=1,...,p$, we solve the following optimization problem:
\begin{equation}\label{lasso:e}
\wh\bb_i=\arg\min_{\bb_i\in R^{m}}\left\{\frac{1}{T}\sum_{t=1}^T(y_{i,t}-\bb_i'\bz_t)^2+\lambda_{i,T}\|\bb_i\|_1\right\},
\end{equation}
where 
$\lambda_{i,T}>0$ is a regularization parameter controlling the sparsity of $\bb_i$ . This approach allows us to efficiently estimate 
$\bB$ and select relevant variables when both 
$m$ and 
$p$ are large, while avoiding overfitting due to the high dimensionality of the regressors 
$\bz_t$. The resulting estimates 
$\wh\bb_i$
  can then be used to compute the residuals 
$\wh\eta_{i,t}=y_{i,t}-\wh\bb_i'\bz_t$, which will be analyzed for further dynamic structure.

Turn to the estimations of $\mathcal{M}(\bA_1)$ and the common factor $\bx_t$. We employ the residuals $\wh\bfeta_t=\by_t-\wh\bB\bz_t$, which represents the part of 
$\by_t$
  that is unexplained by the regression with 
$\bz_t$.
To begin, we introduce some necessary notation first. Let $\bU_1$ and $\bU_2$ be orthonormal complements of $\bA_1$ and $\bA_2$, respectively, that is, $\bU_1\in R^{p\times v}$ and $\bU_2\in R^{\times r}$ are two semi-orthogonal matrices satisfying $\bU_1'\bA_1={\bf 0}$ and $\bU_2'\bA_2={\bf 0}$.  Then $[\bA_1,\bU_1]:=[\ba_1,...,\ba_r,\bu_1,...,\bu_v]$ and $[\bA_2,\bU_2]:=[\ba_{r+1},...,\ba_p,\bu_{v+1},...,\bu_{p}]$ are two $p\times p$ orthonormal matrices.  For $k\geq 1$, let
\[\bSigma_\eta(k)=\Cov(\bfeta_t,\bfeta_{t-k}),\bSigma_x(k)=\Cov(\bx_t,\bx_{t-k}),\,\,\text{and}\,\,\bSigma_{xe}(k)=\Cov(\bx_t,\be_{t-k}),\]
represent the auto-covariance matrices of interest, and let $\bSigma_\eta$, $\bSigma_x$ and $\bSigma_e$ be the covariances of $\bfeta_t$, $\bx_t$ and $\be_t$, respectively.   It follows from (\ref{reg})--(\ref{full}) that
\begin{equation}\label{sigma:eta}
\bSigma_\eta(k)=\bA_1\bSigma_x(k)\bA_1'+\bA_1\bSigma_{xe}(k)\bA_2',\quad k\geq 1,
\end{equation}
and, for $k=0$,
\begin{equation}\label{sigma:eta0}
\bSigma_\eta\equiv \bSigma_\eta(0)=\bA_1\bSigma_x\bA_1'+\bA_2\bSigma_e\bA_2'.
\end{equation}
Next, for a prescribed integer $k_0>0$, define
\begin{equation}\label{M}
\bM=\sum_{k=1}^{k_0}\bSigma_\eta(k)\bSigma_\eta(k)'=\bA_1[\bSigma_x(k)\bA_1'+\bSigma_{xe}(k)\bA_2][\bA_1\bSigma_x(k)'+\bA_2\bSigma_{xe}(k)']\bA_1',
\end{equation}
which is a $p\times p$ semi-positive definite matrix.  Since $\bU_1'\bA_1={\bf 0}$, it follows that $\bM\bU_1={\bf 0}$. This implies that the columns of $\bU_1$ are the eigenvectors corresponding to the zero eigenvalues of $\bM$, while the factor loading space $\mathcal{M}(\bA_1)$ is spanned by the eigenvectors associated with the 
$r$ nonzero eigenvalues of 
$\bM$. Using an approach similar to the orthonormal projections in \cite{gaotsay2021c}, the 
$i$-th column of 
$\bA_1$, denoted as 
$\ba_i$, can be estimated by solving the following optimization problem
\[\max_{\ba}\|\Cov(\ba'\bfeta_t,\bP_t)\|_2^2\quad\text{subject to}\,\, \ba'\ba=1,\]
where $\bP_t:=(\bfeta_{t-1}',...,\bfeta_{t-k_0}')'$ is the pasted lagged vector of $\bfeta_t$. This formulation highlights that the factors recovered by our method exhibit dynamic dependence, as the estimation relies on the relationship between 
$\bfeta_t$
  and its past lagged values. 

To estimate the common factors, we apply the projected PCA method of  \cite{gaotsay2018b}. From Equation (\ref{full}), we have
\begin{equation}\label{u1u2}
\bU_1'\bfeta_t=\bU_1'\bA_2\be_t\quad\text{and}\quad\bU_2'\bfeta_t=\bU_2'\bA_1\bx_t,
\end{equation}
which represent two uncorrelated terms, that is, $\bU_2'\bSigma_\eta\bU_1\bU_1'\bSigma_\eta\bU_2={\bf 0}$.  Here, $\bU_1$ is obtained from the eigen-analysis of $\bM$ in (\ref{M}),  and $\bU_2$ consists of the last $r$ eigenvectors associated with the zero eigenvalues of 
\begin{equation}\label{SS}
   \bS:=\bSigma_\eta\bU_1\bU_1'\bSigma_\eta.
\end{equation}
As discussed in \cite{gaotsay2018b}, $\bU_2'\bA_1$ is invertible. Therefore, from Equation (\ref{u1u2}), we obtain
\[\bx_t=(\bU_2'\bA_1)^{-1}\bU_2'\bfeta_t.\]
This allows us to recover the common factor process 
$\bx_t$
  from the observed residuals 
$\bfeta_t$, utilizing the proposed projected PCA. This method leverages the orthogonality between the signal and noise components and ensures the factors are estimated consistently even in the presence of diverging noise effects.

In practice, given the residuals $\{\wh\bfeta_t| t=1,...,T\}$,  we define the lag-
$k$ sample covariance matrix of 
$\wh\bfeta_t $ as follows:
\[\wh\bSigma_\eta(k)=\frac{1}{T}\sum_{t=k+1}^T(\wh\bfeta_t-\bar{\bfeta})(\wh\bfeta_{t-k}-\bar{\bfeta})',\quad\text{where}\quad\bar{\bfeta}=\frac{1}{T}\sum_{t=1}^T\wh\bfeta_t,\]
and $\wh\bSigma_\eta:=\wh\bSigma_\eta(0)$.  To estimate the factor loading space $\mathcal{M}(\bA_1)$, we perform an eigen-analysis on 
\begin{equation}\label{Mhat}
\wh\bM=\sum_{k=1}^{k_0}\wh\bSigma_\eta(k)\wh\bSigma_\eta(k)',
\end{equation}
where $k_0$ is a prescribed positive integer, as in Equation (\ref{M}). This eigen-analysis provides the eigenvectors that correspond to the nonzero eigenvalues, which span the estimated factor loading space. Let $\wh\bA_1=[\wh\ba_1,...,\wh\ba_r]$ and $\wh\bU_1=[\wh\bu_1,...,\wh\bu_v]$ be the matrices consisting of the eigenvectors of $\wh\bM$ corresponding to the nonzero and zero eigenvalues, respectively., then $\mathcal{M}(\wh\bA_1)$ is the estimator of $\mathcal{M}(\bA_1)$.  In light of the previous discussion, we now proceed to perform an additional eigen-analysis on the matrix
\begin{equation}\label{shat}
\wh\bS=\wh\bSigma_\eta\wh\bU_1\wh\bU_1'\wh\bSigma_\eta,
\end{equation}
as outlined in Equation (\ref{SS}). This represents a sample version 
of the projected PCA approach. 
 Note that $\wh\bS\in R^{p\times p}$ and its rank is at most $p-r$.  As discussed above, we can simply take $\wh\bU_2=[\wh\bu_{v+1},...,\wh\bu_p]$, where the columns vectors are the eigenvectors corresponding to the $r$ smallest eigenvalues of $\wh\bS$.  This approach yields a consistent estimator when the dimension 
$p$ is small. However, when 
$p$ is large, we suggest estimating the number of diverging noise components 
$s$ first, using the following criterion:
\begin{equation}\label{s:est}
\wh s=\arg\min_{1\leq j\leq d_u}{\wh\mu_{j+1}/\wh\mu_{j}},
\end{equation}
where $\wh\mu_1\geq...\geq\wh\mu_p$ are the sample eigenvalues of $\wh\bS$, and $d_u$ is a prescribed positive integer.  Let $\wh\bU_2^*$ be the matrix consisting of the $p-\wh s$ eigenvectors corresponding to the $p-\wh s$ smallest eigenvalues of $\wh\bS$.  We then estimate $\bU_2$ by $\wh\bU_2=\wh\bU_2^*\wh\bR$, where $\wh \bR=[\wh \br_1,...,\wh\br_r]\in R^{(p-\wh d)\times r}$ with $\wh \br_i$ being the eigenvector associated the $i$th largest eigenvalue of $\wh\bU_2^*\wh\bA_1\wh\bA_1'\wh\bU_2^*$. That is, $\wh\bU_2$ is a subspace of $\wh\bU_2^*$. This choice of estimator ensures that the matrix $(\wh\bU_2'\wh\bA_1)^{-1}$ behaves well in recovering the common factors $\wh\bx_t$. Although $\wh\bU_2$ is still not a consistent estimator for $\bU_2$,  it remains orthogonal to the diverging directions of the noises, which is sufficient to mitigate the diverging effect of $\bSigma_e$. Finally, we recover the factor process as $\wh\bx_t=(\wh\bU_2'\wh\bA_1)^{-1}\wh\bU_2'\wh\bfeta_t$.

\subsubsection{Estimation of the Number of Latent Factors}\label{sec2.2.2}


There are several methods available in the literature to determine the number of factors 
$r$ for traditional factor models. For instance, 
{\cite{BaiNg_Econometrica_2002} provides information criteria to 
solve the problem. \cite{lamyao2012} and \cite{ahn2013} consider eigenvalue 
ratio methods and \cite{onatski2010} employs a method based on the random 
matrix theory.

However, when the noise effect is substantial, these traditional methods may not be applicable. \cite{gaotsay2018b} proposes a method based on 
the high-dimensional white noise tests to determine the number of factors, and 
the method continues to apply even when the noise effect is large. 
We adopt this method to determine the number of latent factors 
$r$ in Model (\ref{eta}) using the residuals 
$\wh\bfeta_t$ obtained from the first step of the proposed procedure.}

Specifically,  let $\wh\bG$ be the matrix of eigenvectors (in the decreasing order of the eigenvalues) of the sample matrix $\wh\bM$ in (\ref{Mhat}),  i.e.  $\wh\bG=[\wh\bA_1,\wh\bU_1]$ with $\wh\bA_1\in R^{p\times \wh r}$ and $\wh\bU_1\in R^{p\times \wh v}$, where we still use $\wh\bA_1$ and $\wh\bU_1$ with unknown $\wh r$ (and hence $\wh v$) which is to be determined later.  From the first equation in (\ref{u1u2}) that $\bU_1'\bfeta_t$ is a vector white noise process, then the sample version $\wh\bU_1'\wh\bfeta_t$ will most likely behave like a white noise process.  Let $\wh\bu_t=\wh\bG'\wh\bfeta_t=(\wh u_{1t},...,\wh u_{pt})'$ be the transformed series and $\wh\bu_t(i)=(\wh u_{it},...,\wh u_{pt})'$ for $1\leq i\leq p$, consider the null hypothesis
\[H_0(i): \wh \bu_t(i)\,\,\text{is a vector white noise},\]
with type-I error $\alpha$.  We focus on the case when the dimension 
$p$ is large, as the traditional multivariate Ljung-Box statistics are suitable for testing the hypothesis in scenarios with small 
$p$. For high-dimensional settings, we employ the high-dimensional white noise (HDWN) test introduced by \cite{Tsay_2018}. Let $d_i=p-i+1$ be the length of $\wh\bu_t(i)$ and $\wh\bGamma_{u,k}=[\wh\Gamma_{u,k}(j,l)]_{1\leq j,l\leq d_i}$ be the lag-$k$ sample rank auto-correlation matrix of an orthogonalized vector of $\wh\bu_t(i)$, where the orthogonalization can be done via PCA. 
Although the population covariance of $\bfeta_t$ is of full rank, the residuals $\wh\bfeta_{i,t}$ may encounter some rank loss over the time horizon for $1\leq i\leq p$ due to the subtraction of the regression part from the response $\by_{i,t}$, especially when using traditional least squares with small $m$. However, this rank loss is typically manageable, as the sample size 
$T$ often exceeds the dimension  
$p$ and 
$p$ is significantly larger than the number of regressors 
$m$, as discussed in Section 3. Consequently, we may consider the vector $\wh\bu_{t}(m,i)=(\wh u_{it},...,\wh u_{p-m,t})'$ if $m$ is small. In addition, if $p-m\geq T$, we further reduce the dimension of $\wh\bu_t$ by only keeping the first $p^*$ components where where $p=\epsilon T$ with $0<\epsilon<1$ if $p-m\geq T$. These modifications can ensure that the vector $\wh\bu_t$ can be orthogonalized by PCA. See \cite{gaotsay2018b} for details. The test statistic is defined as
\begin{equation}\label{Tsay:test}
T(N)=\max\{\sqrt{T}|\wh\Gamma_{u,k}(k,l)|:1\leq k,l\leq d_i,1\leq k\leq N\},
\end{equation}
which follows asymptotically the standard Gumbel distribution under the 
null hypothesis $H_0(i)$ based on the extreme value theory. The critical values and rejection regions for this test statistic are provided in closed form and can be found in \cite{Tsay_2018} or in Section 2.3 of \cite{gaotsay2018b}. We initiate the sequential testing procedure with 
$i=1$; if the null hypothesis is rejected, we increment
$i$ by one and repeat the testing process. Using this method, we select 
$\wh r=i-1$ since the largest value is 
$i$ for which the 
$i$-th test fails to reject the null hypothesis.

\subsection{Prediction}

Using $\wh\bB$, $\wh\bA_1$ and $\wh\bx_t$,  we can compute the $h$-step ahead prediction as $\wh\by_{T+h}=\wh\bB\wh\bz_{T+h}+\wh\bA_1\wh\bx_{T+h}$,  where $\wh\bz_{T+h}$ and $\wh\bx_{T+h}$ are  the $h$-step ahead forecast for $\bz_{T+h}$ and $\bx_{T+h}$, respectively,  based on the past observations $\{\bz_T,\bz_{T-1},...,\bz_1\}$ and the estimated  values $\{\wh\bx_T,\wh\bx_{T-1},...,\wh\bx_1\}$. This can be achieved by fitting a VAR or a sparse VAR model to $\{\wh\bz_t,\wh\bz_{t-1},...,\wh\bz_1\}$, depending on the dimension of $\bz_t$, and a VAR model to $\{\wh\bx_t,\wh\bx_{t-1},...,\wh\bx_1\}$, respectively.  Notably, this framework can be extended to the case when $m/p=O(1)$. A specific instance occurs when $\bz_t=(\by_{t-1}',...,\by_{t-q}')'$, containing past lagged variables. In this case, the 1-step ahead forecast is given by $\wh\by_{T+1}=\wh\bB\bz_{T+1}+\wh\bA_1\wh\bx_{T+1}$,  where $\wh\bB\bz_{T+1}$ is observable at time $T$ and $\wh\bx_{T+1}$ is the $1$-step ahead forecast for $\wh\bx_T$ based on the estimated past values $\{\wh\bx_T,\wh\bx_{T-1},...,\wh\bx_1\}$. The prediction for 
$\by_{T+h}$	
  can then be recursively updated using the past predicted values $\wh\by_{T+h-1},...,\wh\by_{T+1}$ and the forecast $\wh\bx_{T+h}$. We omit the details here.

\section{Theoretical Properties}\label{sec3}

We present some asymptotic theory for the estimation method described in Section 2 when $T,p\rightarrow\infty$. We assume $\{\by_t,\bz_t,\bff_t\}$ is $\alpha$-mixing with the mixing coefficients defined by
\begin{equation}\label{amix}
\alpha_p(k)=\sup_{i}\sup_{A\in\mathcal{F}_{-\infty}^i,B\in \mathcal{F}_{i+k}^\infty}|P(A\cap B)-P(A)P(B)|,
\end{equation}
where $\mathcal{F}_i^j$ is the $\sigma$-field generated by $\{(\by_t,\bff_t):i\leq t\leq j\}$.  Let $S_i$ be a subset of $\{1,2,...,m\}$ with cardinality $s_i$ consisting of the indexes of the non-zero components in $\bb_i$, and $S_i^c$ be its complement. Define
\begin{equation}\label{cs}
C_\alpha(S)=\{\Delta\in R^m:\|\Delta_{S^c}\|_1\leq \alpha\|\Delta_S\|_1\}.
\end{equation}
\begin{assumption}\label{a1}
The process $\{\by_t,\bz_t,\bff_t\}$ is $\alpha$-mixing with the mixing coefficients satisfying the condition $\alpha_p(k)<\exp(-k^{\gamma_1})$ for some $\gamma_1>0$, where $\alpha_p(k)$ is defined in (\ref{amix}).
\end{assumption}
\begin{assumption}\label{a2}
For any $1\leq i\leq m$,  $1\leq j\leq r$, $1\leq k\leq v$,  and $1\leq l\leq p$,  $P(|z_{i,t}|>x)\leq C_0\exp(-C_1x^{\gamma_2})$,  $P(|f_{j,t}|>x)\leq C_0\exp(-C_1x^{\gamma_2})$, $P(|\ve_{i,t}|>x)\leq C_0\exp(-C_1x^{\gamma_2})$,  and $P(|\zeta_{i,t}|>x)\leq C_0\exp(-C_1x^{\gamma_2})$ for some $\gamma_2>0$, where $C_0,C_1>0$ is a constant,  $\zeta_{l,t}$ is the $l$-th component of $\bL_2\bve_t$.
\end{assumption}
\begin{assumption}\label{a3}
(i) When $m$ is finite, there exists a constant $C_2>0$ such that $\lambda_{\min}\{E(\bz_t\bz_t')\}>C_2$ for all $t$; (ii) When $m$ is diverging, the matrix $\bZ:=(\bz_1,...,\bz_T)'$ satisfies the restricted eigenvalues condition 
\[\frac{1}{T}\|\bZ\Delta\|_2^2\geq\kappa\|\Delta\|_2^2,\]
for all $\Delta\in C_3(S_i)$ in  (\ref{cs}).
\end{assumption}
Let $\bL=[\bfc_1,...,\bfc_p]$, where $\bfc_i$ is a $p$-dimensional column vector, and $\bL_1=[\bfc_1,...,\bfc_r]$ and $\bL_2=[\bfc_{r+1},...,\bfc_p]$.

\begin{assumption}\label{a4}
There exists a constant $\delta_1\in[0,1)$ such that $\|\bL_1\|_{\min}\asymp\|\bL_1\|_2\asymp p^{(1-\delta_1)/2}$, as the dimension $p$ goes to infinity and $r$ is fixed.
\end{assumption}

\begin{assumption}\label{a5}
$\bL_2$ admits a singular value decomposition $\bL_2=\bA_2\bD_2\bV_2'$, where $\bA_2\in R^{p\times v}$ is given in Equation (\ref{full}), $\bD_2=\diag(d_1,...,d_v)$ and $\bV_2\in R^{v\times v}$ satisfying $\bV_2'\bV_2=\bI_v$, and there exists a finite integer $0< s< v$ such that $d_1\asymp...\asymp d_s\asymp p^{(1-\delta_2)/2}$ for some $\delta_2\in[0,1)$ and $d_{s+1}\asymp...\asymp d_v\asymp 1$.
\end{assumption}


\begin{assumption}\label{a6}
(i) For $\gamma_1$ given in Assumption 1, any $\bh\in R^{v}$ and $0<c_h<\infty$ with $\|\bh\|_2=c_h$, $E|\bh'\bve_t|^{2\gamma_1}<\infty$; (ii)  $\sigma_{\min}(\bR'\bU_2^{*}{'}\bA_1)\geq C_3$ for some constant $C_3>0$ and some half orthogonal matrix $\bR\in R^{(p-s)\times r}$ satisfying $\bR'\bR=\bI_r$, where $\sigma_{\min}$ denotes the minimum non-zero singular value of a matrix. 
\end{assumption}


Assumption \ref{a1}, which addresses dependent random sequences, is standard and well-supported in the literature. For a theoretical justification in the context of VAR models, see \cite{gaoetal2017}. Assumption \ref{a2} guarantees the condition 
$P(|\eta_{i,t}|>x)\leq C\exp(-Cx^{\gamma_2})$, which controls the tail behavior of the noise term 
$\eta_{i,t}$, ensuring sub-exponential decay. Finally, Assumption \ref{a3}(i) ensures that the regressors have a non-singular covariance matrix, allowing the least-squares estimators to be well-defined when 
$m$ is finite. Assumption \ref{a3}(ii) corresponds to the well-known restricted eigenvalue condition used in Lasso regressions; see Chapter 6 of \cite{buhlmann2011}. Alternatively, this condition can be replaced by the more general Restricted Strong Convexity (RSC) condition, which is frequently applied in high-dimensional regularized estimation problems. For further details on RSC, we refer to Chapter 9 of \cite{Wainwright2019}. The parameter 
$\delta_1$ in Assumption \ref{a4} quantifies the strength of the factors, with the eigenvalues of 
$\bL_1\bL_1'$
  being of order 
$p^{1-\delta_1}$. When 
$\delta_1=0$, the factors are considered strong, as this includes the case where each element of 
$\bfc_i$ is $O(1)$. In contrast, if 
$\delta_1>0$, the factors are classified as weak, with smaller values of 
$\delta_1$ corresponding to stronger factors. The advantage of using 
$\delta_1$
  is that it explicitly links the convergence rates of the estimated factors to their strength. Assumptions \ref{a4} and \ref{a5} are similar to those in \cite{gaotsay2018b}, primarily quantifying factor strength and the diverging eigenvalues of the noise terms. Similar assumptions are also made in \cite{lamyao2012}.
Assumption \ref{a6}(i) is mild and includes the  standard normal distribution as a special case. Together with Assumption \ref{a2} and the aforementioned sufficient condition for Assumption \ref{a5}, it is not hard to show $E|y_{it}|^{2\gamma_1}<\infty$, but we do not address this issue explicitly here. 
Assumption \ref{a6}(ii) is reasonable since $\bB_2$ is a subspace of $\bU_2^*$, and 
the invertibility of $\bR'\bU_2^{*}{'}\bA_1$ is illustrated in Remark 1 of \cite{gaotsay2018b}. The choice of $\wh\bR$, and hence $\wh\bU_2=\wh{\bU}_2^*\wh\bR$, will be discussed later.

The following theorem establishes the convergence rates of the estimated regression coefficients for both finite and diverging numbers of regressors.

\begin{theorem}\label{tm1}
(i) Let Assumptions \ref{a1}--\ref{a2} and Assumption \ref{a3}(i) hold.  If $m$ is finite,  as $T,p\rightarrow\infty$, it holds that 
\[\|\wh\bB-\bB\|_F=O_p(p^{1/2}T^{-1/2}).\]
(ii) Let Assumptions \ref{a1}--\ref{a2} and Assumption \ref{a3}(ii) hold.  If $m$ is diverging,  as $T,p\rightarrow\infty$, it holds that 
\[\|\wh\bB-\bB\|_F=O_p(\lambda_T\sqrt{ps^*}),\]
where $\lambda_T\asymp \sqrt{\frac{\log(pm)}{T}}$ and $s^*=\max\{s_i,i=1,...,m\}$.
\end{theorem}
\begin{remark}
(i) From Theorem 1, we see that the stochastic bound between the estimated coefficient matrix and the true one is standard $\sqrt{T}$ when the dimension of the common regressors is finite. On the other hand, if the number of regressors is diverging, the stochastic bound becomes $\sqrt{\frac{p\log(pm)}{T}}$ if the number of nonzero elements in each row of $\bB$ is finite, which is only slightly larger than the one in Theorem 1(i).\\
(ii) Under the assumption that $\lambda_T\sqrt{ps^*}=o(1)$ and additional regularity conditions on the magnitude of the nonzero elements of $\bB$, the sparsity recovery of $\bB$ can be established too. We omit the details here to save space.
\end{remark}


If cross-correlations exist between $\bff_t$ and $\bve_{t-j}$ for $j\geq 1$, we assume rank($\bSigma_{f\epsilon}(k)$)$=r$ and define
\begin{equation}\label{kmm}
\kappa_{\min}=\min_{1\leq k\leq k_0}\|\bSigma_{f\epsilon}(k)\|_{\min}\,\, \text{and}\,\, \kappa_{\max}=\max_{1\leq k\leq k_0}\|\bSigma_{f\epsilon}(k)\|_2,
\end{equation}
where $\|\cdot\|_{\min}$ denotes the smallest nonzero singular value, $\kappa_{\min}$ and $\kappa_{\max}$ can be either finite constants or diverging rates in relation to $p$ and $T$, and they control the strength of the dependence between $\bff_t$ and the past errors $\bve_{t-j}$ for $j\geq 1$. The maximal order of $\kappa_{\max}$ is $p^{1/2}$ which is the Frobenius norm of $\bSigma_{f\ve}(k)$, and $\kappa_{\max}=0$ (hence $\kappa_{\min}=0$) implies that $\bff_t$ and $\bve_s$ are uncorrelated for all $t$ and $s$.
Throughout this article, if $\bff_t$ and $\bve_s$ are independent for all $t$ and $s$ which is stronger than being uncorrelated, then $\kappa_{\min}=\kappa_{\max}=0$ and all the conditions and expressions below concerning  $\kappa_{\min}$ and $\kappa_{\max}$ are removed.

To this end, we adopt the discrepancy measure used by
\cite{panyao2008}: for two $p\times r$ half orthogonal
matrices ${\bf H}_1$ and ${\bf H}_2$ satisfying the condition ${\bf
H}_1'{\bf H}_1={\bf H}_2'{\bf H}_2=\bI_{r}$, the difference
between the two linear spaces $\mathcal{M}({\bf H}_1)$ and
$\mathcal{M}({\bf H}_2)$ is measured by
\begin{equation}
D({\bf H}_1,{\bf
H}_2)=\sqrt{1-\frac{1}{r}\textrm{tr}({\bf H}_1{\bf H}_1'{\bf
H}_2{\bf H}_2')}.\label{eq:D}
\end{equation}
Note that $D({\bf H}_1,{\bf H}_2) \in [0,1].$ 
It is equal to $0$ if and only if
$\mathcal{M}({\bf H}_1)=\mathcal{M}({\bf H}_2)$, and to $1$ if and
only if $\mathcal{M}({\bf H}_1)\perp \mathcal{M}({\bf H}_2)$. The distance defined in \ref{eq:D} is equivalent to the so called $\sin(\bTheta)$ distance in the literature when $r$ is finite. See Theorem I.5.5 of \cite{stewart-sun1990} or \cite{gaotsay2021c} for illustrations.

The following theorem establishes the convergence of $D(\widehat{\bf A}_1,{\bf A}_1)$ when $\bA_1$ is not uniquely defined.
The following theorem establishes the consistency of the estimated loading matrices.

\begin{theorem}\label{tm3}
Suppose Assumptions \ref{a1}--\ref{a6} hold and $r$ is known and fixed. When $m$ is finite,  assume $p^{\delta_1/2}T^{1/2}=o(1)$ and $p^{\delta_2/2}T^{1/2}=o(1)$; when $m$ is diverging, the conditions $p^{\delta_1/2}\log(pm)s^*T^{-1/2}=o(1)$ $p^{\delta_1/2}s^*\log(pm)T^{-1/2}=o(1)$ hold.  
(i) Under the condition that $\delta_1\leq \delta_2$,
\[D(\wh\bA_1,\bA_1)=\left\{\begin{array}{ll}
O_p(p^{\delta_1/2}T^{-1/2}),&\text{if}\,\, \kappa_{\max}p^{\delta_1/2-\delta_2/2}=o(1),\\
O_p(\kappa_{\min}^{-2}p^{\delta_2-\delta_1/2}T^{-1/2}+\kappa_{\min}^{-2}\kappa_{\max}p^{\delta_2/2}T^{-1/2}),&\text{if}\,\, r\leq s, \kappa_{\min}^{-1}p^{\delta_2/2-\delta_1/2}=o(1),\\
O_p(\kappa_{\min}^{-2}p^{1-\delta_1/2}T^{-1/2}+\kappa_{\min}^{-2}\kappa_{\max}p^{1-\delta_2/2}T^{-1/2}),&\text{if}\,\, r>s, \kappa_{\min}^{-1}p^{(1-\delta_1)/2}=o(1),
\end{array}\right.\]
and the above results also hold for $D(\wh\bU_1,\bU_1)$, and 
\[D(\wh{\bU}_2^*,\bU_2^*)=O_p(p^{2\delta_2-3\delta_1/2}T^{-1/2}+p^{2\delta_2-2\delta_1}D(\wh\bU_1,\bU_1)).\]

(ii) Under the condition that $\delta_1>\delta_2$, if $\kappa_{\max}=0$ and $p^{\delta_1-\delta_2/2}T^{-1/2}=o(1)$, then
\[D(\wh\bA_1,\bA_1)=O_p(p^{\delta_1-\delta_2/2}T^{-1/2}).\]
If $\kappa_{\max}\geq\kappa_{\min}\geq c>0$ for some constant $c$,
then
\[D(\wh\bA_1,\bA_1)=\left\{\begin{array}{ll}
O_p(\kappa_{\min}^{-2}\kappa_{\max}p^{\delta_1/2}T^{-1/2}),&\text{if}\,\, r\leq K, \kappa_{\min}^{-1}p^{\delta_2/2-\delta_1/2}=o(1),\\
O_p(\kappa_{\min}^{-2}\kappa_{\max}p^{1+\delta_1/2-\delta_2}T^{-1/2}),&\text{if}\,\, r>K, \kappa_{\min}^{-1}p^{(1-\delta_1)/2}=o(1),
\end{array}\right.\]
and the above results also hold for $D(\wh\bU_1,\bU_1)$, and
\[D(\wh{\bU}_2^*,\bU_2^*)=O_p(p^{\delta_2/2}T^{-1/2}+D(\wh\bU_1,\bU_1)).\]

\end{theorem}
\begin{remark}

The consistency of all estimated parameters is discussed in \cite{gaotsay2018b} under various scenarios for $p$ and $T$. When $\kappa_{max}=\kappa_{\min}=0$, i.e., the factor process $\bx_t$ are uncorrelated with all the past white noises, we have
\[D(\wh\bA_1,\bA_1)=O_p(p^{\delta_1/2}T^{-1/2})\,\,\text{and}\,\,D(\wh{\bU}_2^*,\bU_2^*)=O_p(p^{2\delta_2-3\delta_1/2}T^{-1/2}),\,\,\text{if}\,\,\delta_1\leq \delta_2,\]
and
\[D(\wh\bA_1,\bA_1)=O_p(p^{\delta_1-\delta_2/2}T^{-1/2})\,\,\text{and}\,\,D(\wh{\bU}_2^*,\bU_2^*)=O_p(p^{\delta_2/2}T^{-1/2}+p^{\delta_1-\delta_2/2}T^{-1/2}),\,\,\text{if}\,\,\delta_1> \delta_2.\]
In particular, when all the factors and the diverging noises are strong, i.e. $\delta_1=\delta_2=0$, all the convergence rates will become standard $\sqrt{T}$.
\end{remark}

We now state the consistency of the estimated factor terms.
\begin{theorem}\label{tm4}
Under the conditions in Theorem \ref{tm3}, we have
\begin{equation*}\label{ex:f}
p^{-1/2}\|\wh\bA_1\wh\bx_t-\bA_1\bx_t\|_2=O_p\{p^{-1/2}+p^{-\delta_1/2}D(\wh\bA_1,\bA_1)+p^{-\delta_2/2}D(\wh\bU_2^*,\bU_2^*)+\sqrt{\frac{ms^*\log(pm)}{T}}\}.
\end{equation*}
\end{theorem}
\begin{remark}
Similarly, we can simplify the stochastic bound in Theorem \ref{tm4} when the dependence between $\bff_t$ and the past white noise $\bve_{t-j}$ for $j\geq 0$ is zero. In particular, when  $\delta_1=\delta_2=0$, that is, the factors and the noise terms are all strong, the stochastic bound in (\ref{tm4}) is $O_p(p^{-1/2}+T^{-1/2}+\sqrt{\frac{ms^*\log(pm)}{T}})$, which has one more term $\sqrt{\frac{ms^*\log(pm)}{T}}$  compared to the rate specified in Theorem 3 of \cite{Bai_Econometrica_2003} when dealing with the traditional approximate factor models. This discrepancy arises from the estimation error associated with the regression coefficient matrix in the first step.
\end{remark}

The following theorem establishes the consistency of the estimated number of latent factors using high-dimensional white-noise test.
\begin{theorem}\label{tm5}
 Suppose Assumptions \ref{a1}--\ref{a6} hold.  If $\sqrt{p}(\log(Tp))^{1/\gamma_2}D(\wh\bU_1,\bU_1)=o_p(1)$ and $\sqrt{\frac{pms^*\log(pm)}{T}}(\log(Tm))^{1/\gamma_2}=o(1)$, then 
the test statistic $T(m)$ of  (\ref{Tsay:test}) can also consistently estimate $r$.
\end{theorem}
\begin{remark}
From Theorem~\ref{tm5}, we see that the conditions can be simplified to $p^{1/2}\log(Tp)T^{-1/2}=o(1)$ and $p^{1/2}\sqrt{\log(p)}\log(T)T^{-1/2}=o(1)$ under the assumptions that $\delta_1=\delta_2=0$, $m$ and $s^*$ are finite, and $\gamma_2=1$. Consequently, if $p/T\rightarrow 0$, the consistency of $\wh r$ can be guaranteed, and all the estimators are consistent according to Theorems \ref{tm1}--\ref{tm4} under such assumptions.
\end{remark}

\section{Numerical Properties}\label{sec4}
In this section, we use simulation and a real example to assess the performance of the proposed procedure in finite samples.

\subsection{Simulation}
As the dimensions of $\wh\bA_1$  and $\bA_1$ are not necessarily the same,
and  $\bL_1$ is not an orthogonal matrix in general, we first extend the discrepancy measure
in Equation (\ref{eq:D}) to a more general form below. Let $\bH_i$ be a
$p\times r_i$ matrix with rank$(\bH_i) = r_i$, and $\bP_i =
\bH_i(\bH_i'\bH_i)^{-1} \bH_i'$, $i=1,2$. Define
\begin{equation}\label{dmeasure}
\bar{D}(\bH_1,\bH_2)=\sqrt{1-
\frac{1}{\max{(r_1,r_2)}}\textrm{tr}(\bP_1\bP_2)}.
\end{equation}
Then $\bar{D} \in [0,1]$. Furthermore,
$\bar{D}(\bH_1,\bH_2)=0$ if and only if
either $\mathcal{M}(\bH_1)\subset \mathcal{M}(\bH_2)$ or
$\mathcal{M}(\bH_2)\subset \mathcal{M}(\bH_1)$, and  it is 1 if and only if
$\mathcal{M}(\bH_1) \perp \mathcal{M}(\bH_2)$.
When $r_1 = r_2=r$ and $\bH_i'\bH_i= \bI_r$,
$\bar{D}(\bH_1,\bH_2)
$ is the same as that in Equation (\ref{eq:D}). We only present the simulation results 
for $k_0=2$ in Equation (\ref{Mhat}) to save space because other choices of $k_0$ 
produce similar patterns according to our simulation studies.\\ 


\noindent{\bf Example 1.}
Consider Models (\ref{reg})--(\ref{eta}) with the regressors and factors  following 
\[\bz_t=\bPhi_1\bz_{t-1}+\bxi_t,\,\,\text{and}\,\,\bff_t=\bPhi_2\bff_{t-1}+\bw_t,\]
respectively, where $\bxi_t$ and $\bw_t$ are two white noise processes. We set the true numbers of regressors to $m=5$, the number of factors to $r=3$, and the spiked components of $\bL_2\bve_t$ to $s=3$, as defined in Assumption 5. The dimensions used are $p=50,100,150,200$, and the sample sizes are $T=300,500,1000,1500$. We consider three scenarios for the strength parameters $\delta_1$ and $\delta_2$ with $(\delta_1,\delta_2)=(0,0),(0.4,0.5)$ and $(0.5,0.4)$, respectively. 
For each realization of $\by_t$ and dimension $p$, we first set the seed number in \texttt{R} to \texttt{1234}, generate a $p\times p$ matrix, and perform a singular value decomposition to obtain a matrix $\bQ$ consisting of its left singular vectors. Partition $\bQ=[\bQ_1,\bQ_2]$, where $\bQ_1\in R^{p\times r}$ and $\bQ_2\in R^{p\times (p-r)}$, $\bL_1$ is chosen as $\bQ_1$ multiplied by $p^{(1-\delta_1)/2}\bI_r$ on its right, and $\bL_2$ is chosen as $\bQ_2$ multiplied by $[p^{(1-\delta_2)/2}\bI_s,2\bI_{p-r-s}]$ on the right. In other words, the eigenvalues of the factors and the spiked components of white noises are $p^{(1-\delta_1)}$ and $p^{(1-\delta_2)}$, respectively, and  the 
remaining nonzero eigenvalues of noises are $4$. $\bPhi_1$ and $\Phi_2$ are diagonal matrices with their diagonal elements being drawn independently from $U(0.5,0.9)$, $\bxi_t\sim N(0,\bI_m)$, $\bw_t\sim N(0,\bI_r)$, and $\bve_t\sim N(0,\bI_v)$. We use 500 replications for each $(p,T)$ configuration.

\begin{figure}
\begin{center}
{\includegraphics[width=0.7\textwidth]{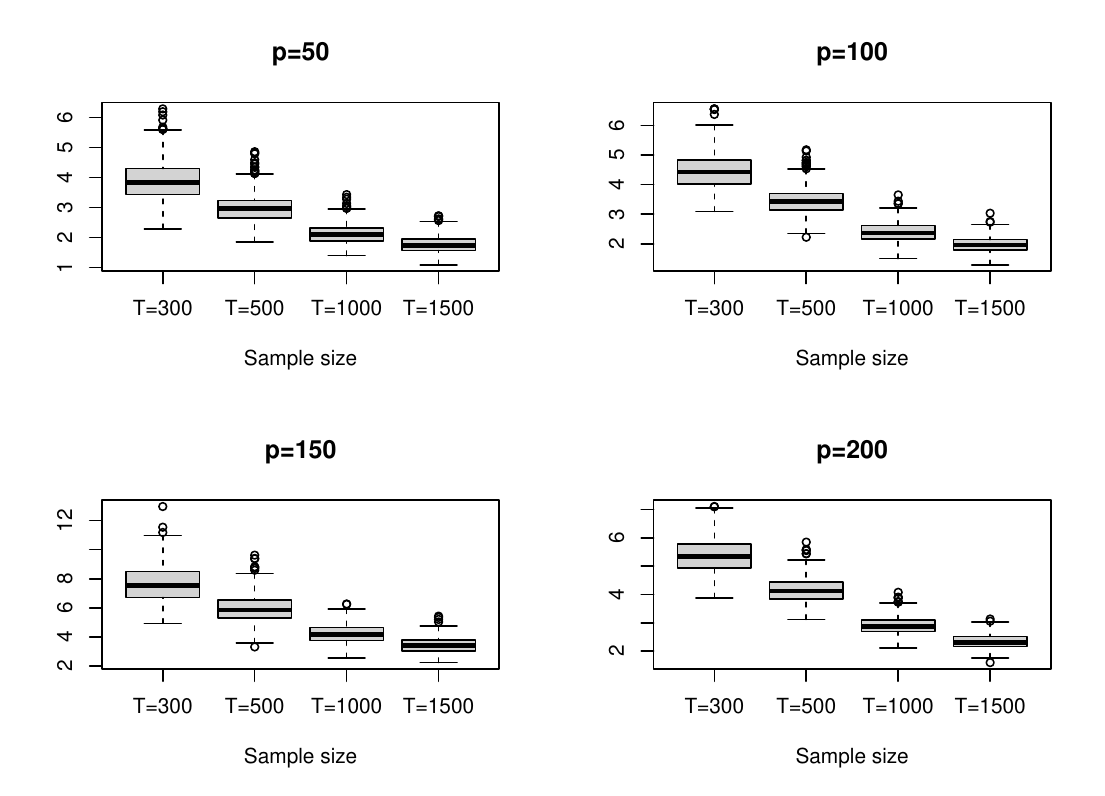}}
\caption{Boxplots of $\|\wh\bB-\bB\|_F$ when $m=3$, $r=3$, and $s=3$ under different scenarios of $p$ in Example 1 with $(\delta_1,\delta_2)=(0,0)$. The sample sizes are $300, 500, 1000$, and $1500$. $500$ iterations are used in the experiment.}\label{fig1}
\end{center}
\end{figure}

We first study the performance of the component-wise regression. Figure~\ref{fig1} plots the Frobenius-norm of the difference between the estimated regression coefficient matrices and the true one when $(\delta_1,\delta_2)=(0,0)$. We obtained similar findings for other choices. From Figure~\ref{fig1}, we see that, for each $p$, the discrepancy decreases as the sample size increases, and this is in agreement with the theory. We also note that the discrepancy may not decrease as the dimension $p$ increases. 
This is so because it also depends on the strength of the error terms. For example, under the current simulation setting, the average variance of the error terms when $p=200$ is even smaller than that of $p=150$, which explains that the Frobenius-norm decreases when $p$ increases from $150$ to $200$.

\begin{figure}
\begin{center}
{\includegraphics[width=0.7\textwidth]{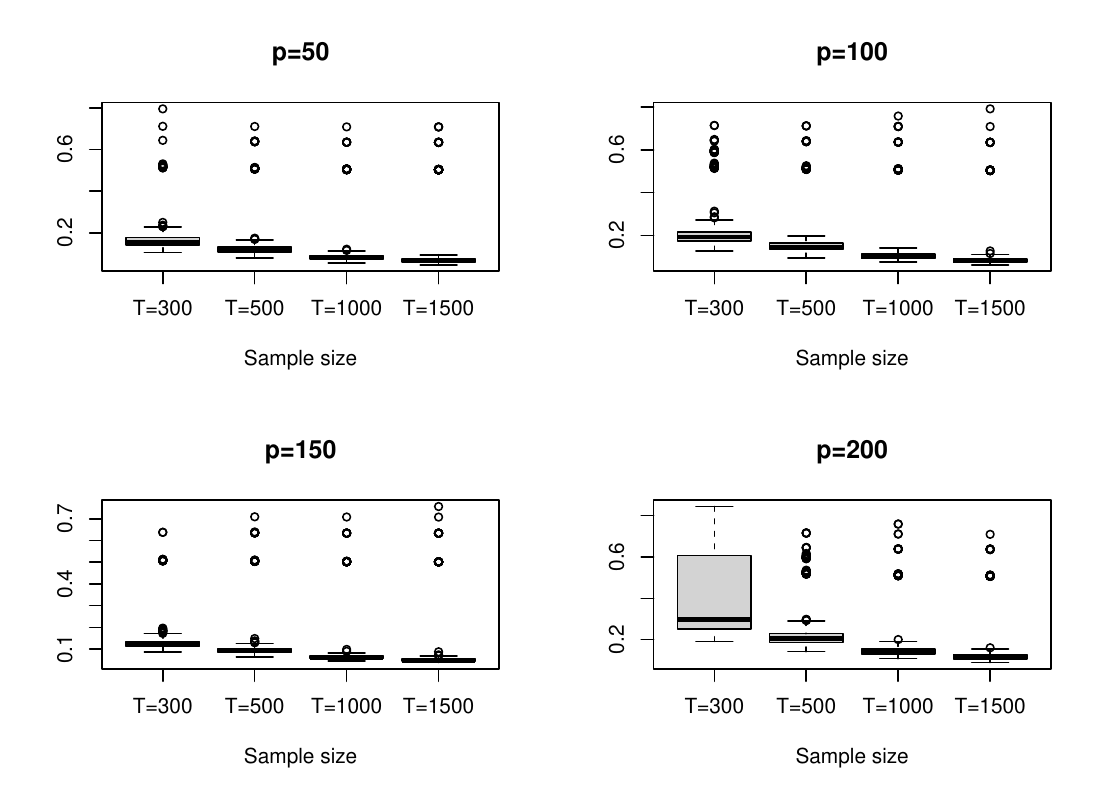}}
\caption{Boxplots of $\bar{D}(\wh\bA_1,\bL_1)$ when $m=3$, $r=3$, and $s=3$ under different scenarios of $p$ in Example 1 with $(\delta_1,\delta_2)=(0,0)$. The sample sizes are $300, 500, 1000, 1500$. $500$ iterations are used  in the experiment.}\label{fig2}
\end{center}
\end{figure}

After removing the regression component, we apply the proposed eigen-analysis to the residuals $\wh\bfeta_t$. We first study the performance of estimating the number of factors using the residuals. Table~\ref{Table1} reports the empirical probabilities $P(\wh r=r)$. From the table, 
we see that, for each setting of $(\delta_1,\delta_2)$, the performance of the white noise test is quite satisfactory for all configurations of $(p,T)$. For each $p$, the performance may not strictly improve as the sample size increases, because the test involves a PCA orthogonalization which is sensitive to the choices of $p$ and $T$. The overall performance of the proposed procedure 
is satisfactory for large $T$ as can be seen in Table~\ref{Table1}. To study the estimated loading matrices, we present the boxplots of $\bar{D}(\wh\bA,\bL_1)$ in Figure~\ref{fig2}.  From Figure~\ref{fig2}, there is a clear pattern that the estimation accuracy of the loading matrix improves as the sample size increases, which is in line with our asymptotic theory. It also shows that the  white noise test selects $\wh r$ reasonably well in finite samples.
\begin{table}
 \caption{Empirical probabilities $P(\hat{r}=r)$ of Example 1 with $m=5$, $r=3$, and $s=3$, where $p$ and $T$ are the dimension and the sample size, respectively. $\delta_1$ and $\delta_2$ are the strength parameters of the factors and the errors, respectively. $500$ iterations are used in the experiment.}
          \label{Table1}
\begin{center}
 \setlength{\abovecaptionskip}{0pt}
\setlength{\belowcaptionskip}{3pt}

\begin{tabular}{c|c|cccc}
\hline
&&\multicolumn{4}{c}{$T$}\\
$(\delta_1,\delta_2)$ &$p$&$300$&$500$&$1000$&$1500$\\
\hline
(0,0)&$50$&0.956&0.916&0.910&0.880\\
&$100$&0.888&0.892&0.912&0.912\\
&$150$&0.950&0.912&0.920&0.926\\
&$200$&0.584&0.870&0.912&0.908\\
\hline
(0.4,0.5)&$50$&0.936&0.898&0.916&0.894\\
&$100$&0.934&0.906&0.926&0.922\\
&$150$&0.938&0.920&0.918&0.914\\
&$200$&0.942&0.930&0.930&0.916\\
   \hline 
  (0.5,0.4)&$50$&0.792&0.910&0.912&0.882\\
&$100$&0.642&0.888&0.910&0.912\\
&$150$&0.906&0.920&0.914&0.926\\
&$200$&0.656&0.922&0.942&0.926\\
\hline
\end{tabular}
          \end{center}
\end{table}

\begin{figure}
\begin{center}
{\includegraphics[width=0.7\textwidth]{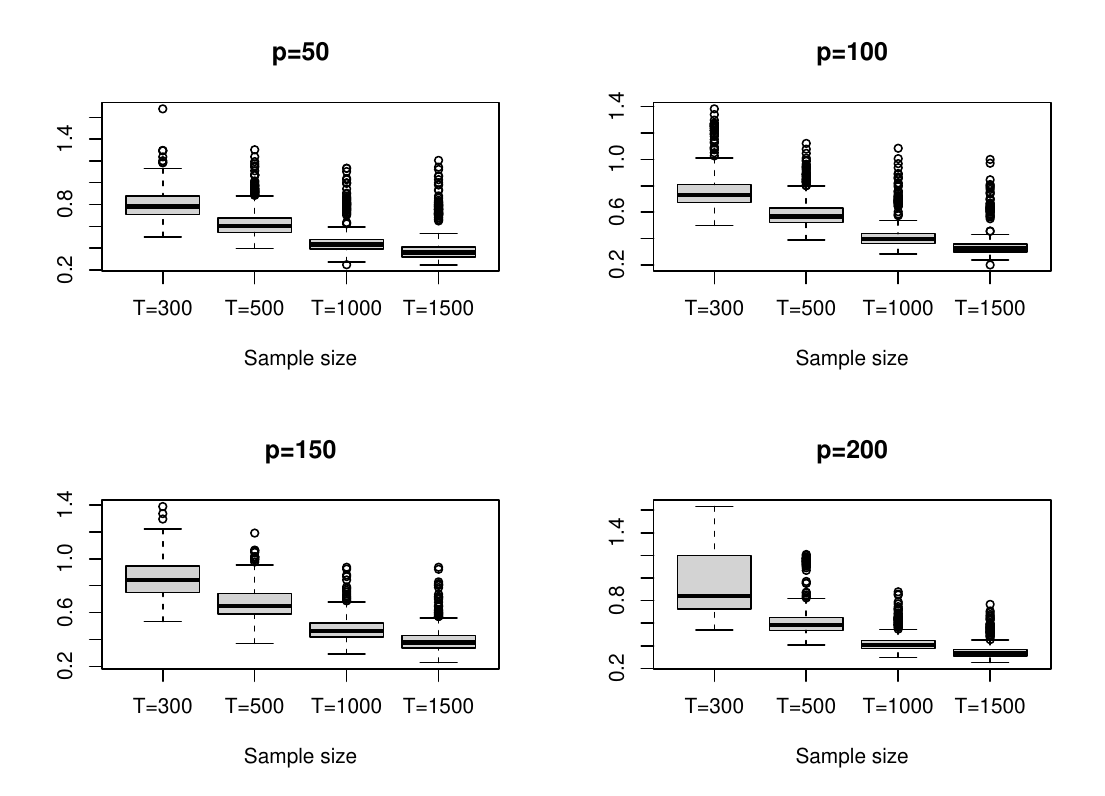}}
\caption{Boxplots of the RMSE defined in (\ref{rmse:plg}) when $m=3$, $r=3$, and $s=3$ under different scenarios of $p$ in Example 1 with $(\delta_1,\delta_2)=(0,0)$. The sample sizes are $300, 500, 1000, 1500$. $500$ iterations are used in the experiment.}\label{fig3}
\end{center}
\end{figure}

To study the the accuracy in estimating the common factors, define the RMSE as
\begin{equation}\label{rmse:plg}
\text{RMSE}=\left(\frac{1}{Tp}\sum_{t=1}^T\|\wh\bA_1\wh\bx_t-\bL_1\bff_t\|_2^2\right)^{1/2},
\end{equation}
which quantifies the estimation accuracy of the common factor process. The results are shown in Figure~\ref{fig3}. The performance is similar as before, and we omit the details.\\

\noindent {\bf Example 2.} In this example, we consider the case when the number of regressors $m$ is large with only a finite number of nonzero elements in each row of the coefficient matrix $\bB$. We set $m=40$ and the sparsity parameters $s_1=...=s_p=5$. The findings are similar for other choices. For each iteration, the seed number used to generate the coefficient matrix and other parameters are the same as those of Example 1. We first draw elements of $\bB$ uniformly from $U(-2,-1)\cup U(1,2)$, then we randomly select $s_i$ elements in the $i$th row of $\bB$ as nonzero parameters, and set others to zero. All the remaining settings are the same as those in Example 1.

\begin{figure}
\begin{center}
{\includegraphics[width=0.7\textwidth]{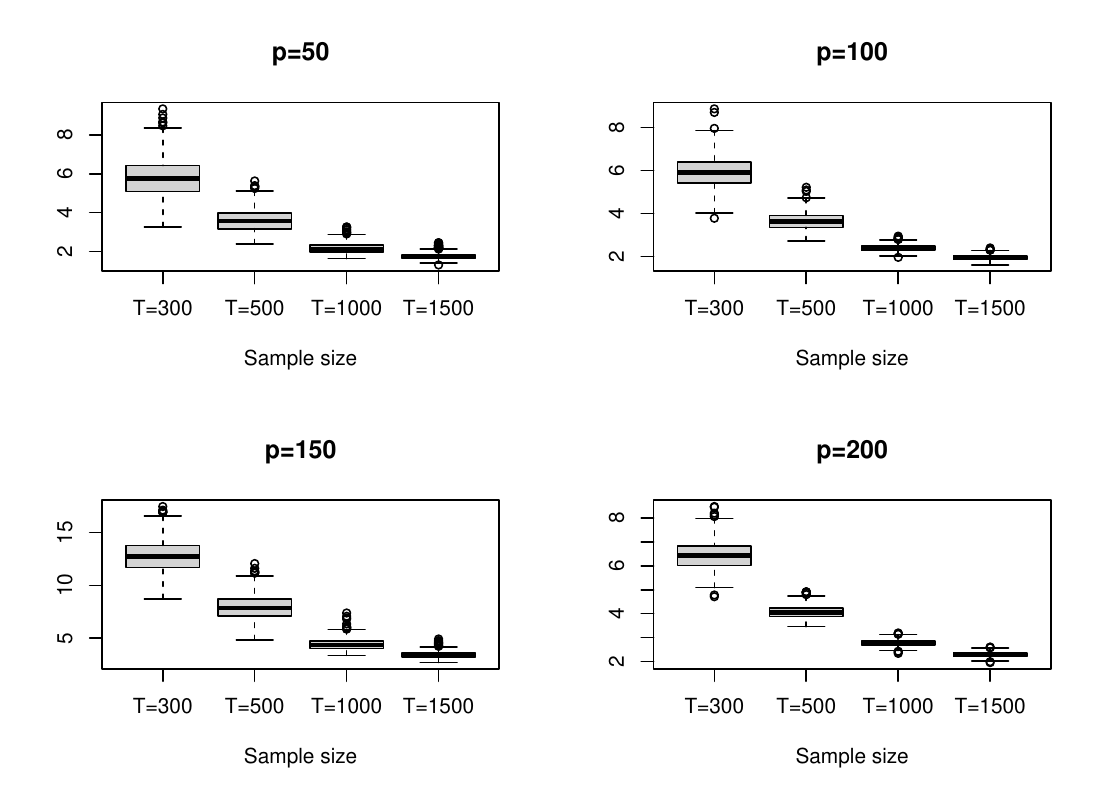}}
\caption{Boxplots of $\|\wh\bB-\bB\|_F$ when $m=40$, $r=3$, $s=3$ and the sparsity parameters $s_i=5$ for $1\leq i\leq p$ under different scenarios of $p$ in Example 2 with $(\delta_1,\delta_2)=(0,0)$. The sample sizes are $300, 500, 1000, 1500$. $500$ iterations are used in the experiment.}\label{fig4}
\end{center}
\end{figure}

We first study the performance of the sparse regression. In each series, we use \texttt{lars} in \texttt{R} and estimate the non-zero elements using Lasso. To obtain more accurate estimates of the residuals, we modify the Lasso approach in the simulation as follows. We first estimate the sparse parameters $s_i$ and obtain the nonzero indexes in the $i$th row, then we estimate the nonzero parameters using only $\wh s_i$ relevant regressors identified by Lasso. Figure~\ref{fig4} presents the Frobenius-norm of the difference between the estimated regression coefficient matrices and the true one when $(\delta_1,\delta_2)=(0,0)$. Similar findings can be obtained for other choices. The 
general finding is the same as before and it is clear that the discrepancy decreases as the sample size increases for each $p$.

We next study the performance of the white noise test when the Lasso procedure is applied to the high-dimensional regression. Table~\ref{Table2} reports the empirical probabilities $P(\wh r=r)$ when applying the white-noise test to the residuals. From Table~\ref{Table2}, we see that for each setting of $(\delta_1,\delta_2)$ and fixed dimension $p$, the performance of the white noise test improves as the sample size increases for most cases. We also note that the power of the test is pretty low when $p=150$ for $(\delta_1,\delta_2)=(0,0)$. This might be due to the larger errors resulted from the first step regression for $p=150$. See the bottom-left plot in Figure~\ref{fig4}. The performance can be improved by changing the setting of the parameters in $\bL_2$. For example, if we choose $\bL_2$  as $\bQ_2$ multiplied by $[p^{(1-\delta_2)/2}\bI_s,3\bI_{p-r-s}]$ on the right, where the bounded nonzero eigenvalues of the noise become $3$ instead of being $2$ as in Table~\ref{Table2}, the empirical probabilities for $p=150$ and $(\delta_1,\delta_2)=(0,0)$ become $(0.038,0.144,0.580,0.798)$ for $T=300, 500,1000,1500$, respectively. We omit the details here.
\begin{table}
 \caption{Empirical probabilities $P(\hat{r}=r)$ of Example 2 with $m=40$, $r=3$ and $s=3$, where $p$ and $T$ are the dimension and the sample size, respectively. The sparsity parameters $s_i=5$ for $1\leq i\leq p$. $\delta_1$ and $\delta_2$ are the strength parameters of the factors and the errors, respectively. $500$ iterations are used in the experiment.}
          \label{Table2}
\begin{center}
 \setlength{\abovecaptionskip}{0pt}
\setlength{\belowcaptionskip}{3pt}

\begin{tabular}{c|c|cccc}
\hline
&&\multicolumn{4}{c}{$T$}\\
$(\delta_1,\delta_2)$ &$p$&$300$&$500$&$1000$&$1500$\\
\hline
(0,0)&$50$&0.136&0.264&0.684&0.808\\
&$100$&0.346&0.702&0.928&0.912\\
&$150$&0&0&0.052&0.166\\
&$200$&0.464&0.878&0.908&0.930\\
\hline
(0.4,0.5)&$50$&0.942&0.916&0.918&0.928\\
&$100$&0.948&0.926&0.912&0.918\\
&$150$&0.948&0.948&0.914&0.918\\
&$200$&0.914&0.942&0.936&0.926\\
   \hline 
  (0.5,0.4)&$50$&0.794&0.918&0.908&0.920\\
&$100$&0.698&0.914&0.916&0.910\\
&$150$&0.876&0.936&0.902&0.908\\
&$200$&0.598&0.948&0.930&0.912\\
\hline
\end{tabular}
          \end{center}
\end{table}

\begin{figure}
\begin{center}
{\includegraphics[width=0.7\textwidth]{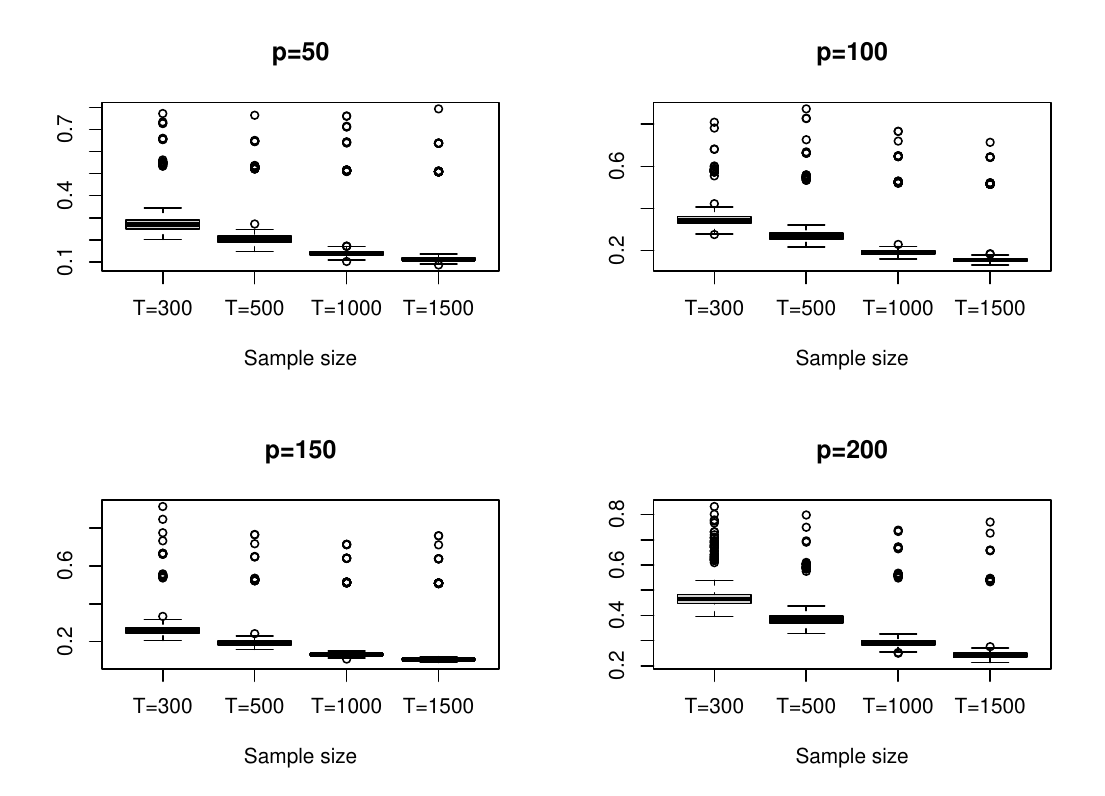}}
\caption{Boxplots of $\bar{D}(\wh\bA_1,\bL_1)$ when $m=40$, $r=3$, $s=3$ and the sparsity parameters $s_i=5$ for $1\leq i\leq p$ under different scenarios of $p$ in Example 2 with $(\delta_1,\delta_2)=(0.4,0.5)$. The sample sizes are $300, 500, 1000, 1500$. $500$ iterations are used in the experiment.}\label{fig5}
\end{center}
\end{figure}

\begin{figure}
\begin{center}
{\includegraphics[width=0.7\textwidth]{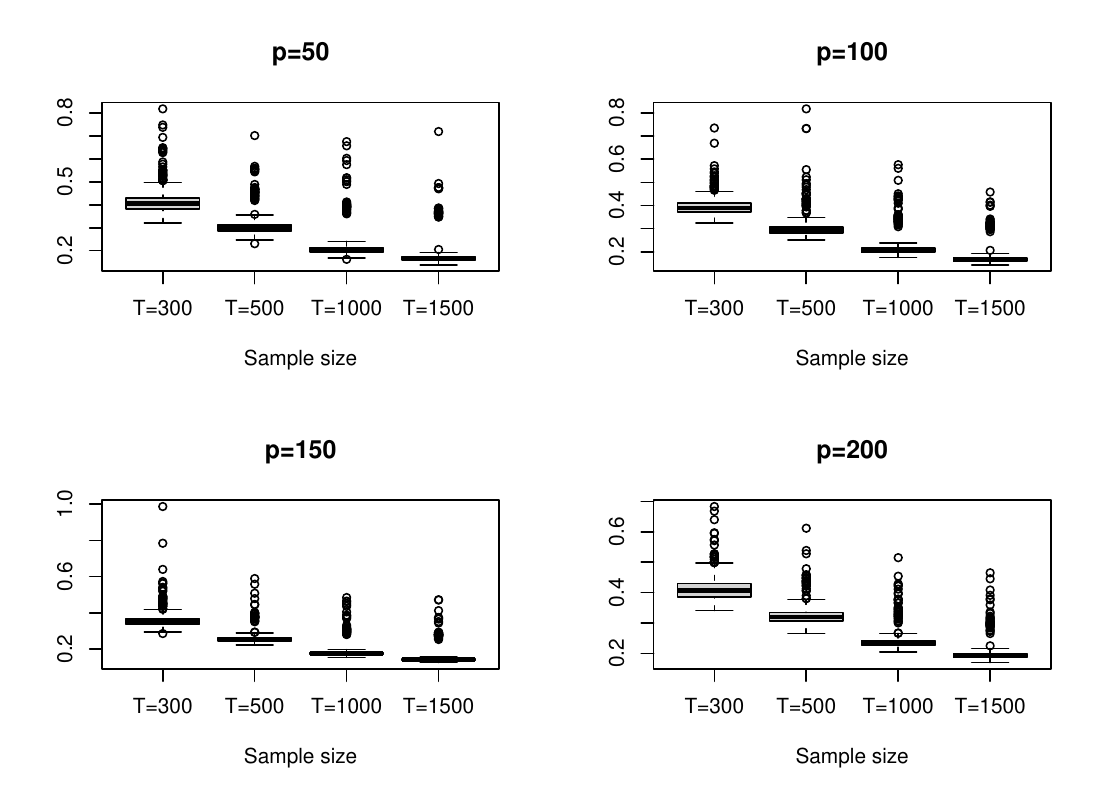}}
\caption{Boxplots of the RMSE defined in (\ref{rmse:plg}) when $m=40$, $r=3$, $s=3$ and the sparsity parameters $s_i=5$ for $1\leq i\leq p$ under different scenarios of $p$ in Example 2 with $(\delta_1,\delta_2)=(0.4,0.5)$. The sample sizes are $300, 500, 1000, 1500$. $500$ iterations are used.}\label{fig6}
\end{center}
\end{figure}

Now we study the estimated loading matrices and the estimation accuracy of the common factors, similar as those in Example 1, the boxplots of $\bar{D}(\wh\bA_1,\bL_1)$ and the RMSE defined in (\ref{rmse:plg}) are shown in Figure~\ref{fig5} and Figure~\ref{fig6}, respectively, for the case when  $(\delta_1,\delta_1)=(0.4,0.5)$. Similar patterns can also be obtained for other choices. There is a clear pattern that the estimation accuracy of the loading matrices and the common factors improve as the sample size increases, which is in line with our asymptotic theory. Overall, the simulations show that 
the proposed procedure is capable of producing satisfactory results.

\subsection{Real Data Analysis}
\begin{figure}
\begin{center}
{\includegraphics[width=0.7\textwidth]{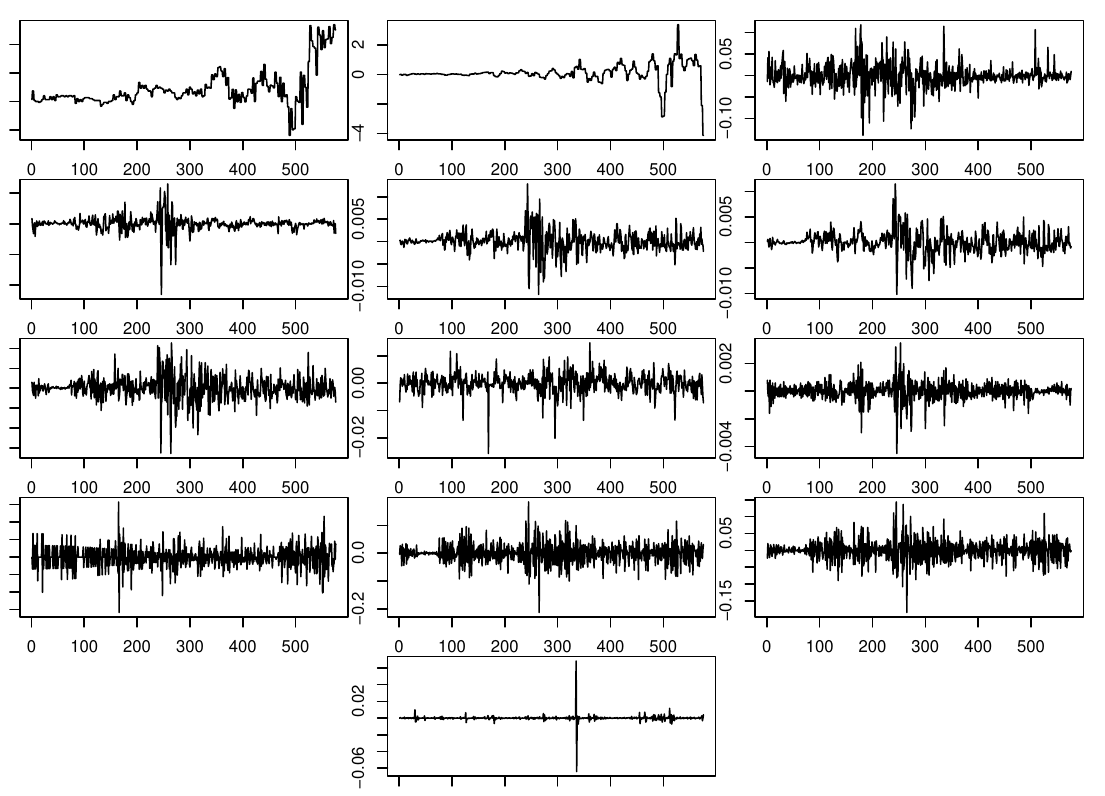}}
\caption{Time series plots of the 13 predictors used in Example 3 during the time from December, 1959 to November, 2007. }\label{fig7}
\end{center}
\end{figure}

In this section, we illustrate the proposed methodology by a dataset consisting of stock returns and some predictors identified in the finance literature.\\
\noindent {\bf Example 3.} Consider a dataset of monthly stock returns, which can be downloaded from the Wharton Research Data Services (WRDS) website \url{https://wrds-www.wharton.upenn.edu/}. We choose the stocks that are included in the S\&P 500 index and find that there are 119 stocks which have 
data available during the time spanning from January, 1960 to December, 2007 without any missing values. Therefore, we have $p=119$ and $T=576$ for the dataset. In addition, the regressors are chosen because they have been found to have predictive ability for the stock returns, as identified in \cite{welch2008}. See an updated version of the data on the website (\url{https://sites.google.com/view/agoyal145}). We eliminate the predictors which have missing values during the time span used and take a first difference to remove any nonstationarity. This results in 13 predictors available, which are shown in Figure~\ref{fig7}, where we use one lag behind the stock return data for the purpose of investigating the predictive performance later.
The descriptions of the predictors are given in Table~\ref{Table3}. Overall, 
we have $m=13$, $p=119$, and $T=576$ in our empirical study.

\begin{table}
 \caption{Definitions of predictors in Figure~\ref{fig7} (by row).}
          \label{Table3}
\begin{center}
 \setlength{\abovecaptionskip}{0pt}
\setlength{\belowcaptionskip}{3pt}

\begin{tabular}{p{1.5 cm}|p{14cm}}
\hline
Predictor&Definition\\
\hline
Dividends&Dividends are 12-month moving sums of dividends paid on the S\&P 500 index.\\
Earnings&Earnings are 12-month moving sums of earnings on the S\&P 500 index.\\
B/M& The book-to-market-ratio is the ratio of book value of market value for the Dow Jones Jones Industrial Average\\
TBL& Treasury bill rates: 3-month treasury bill rates\\
AAA&Corporate Bond Yields on AAA-rated bonds\\
BAA&Corporate Bond Yields on BAA-rated bonds\\
LTY&Long-term yield: long-term government bond yield\\
NTIS&Net equity expansion: ratio of 12-month moving sums of net issues by NYSE listed stocks over the total end-of year market capitalization of NYSE stocks\\
RFree&Risk free month rate\\
INFL&Inflation: CPI inflation for all urban consumers\\
LTR&Long-term return: return of long term government bounds\\
CORPR&Long-term corporate bond returns\\
SVAR&Stock variance: sum of squared daily returns on S\&P 500 index\\
\hline
\end{tabular}
          \end{center}
\end{table}
 
\begin{figure}
\begin{center}
{\includegraphics[width=0.8\textwidth]{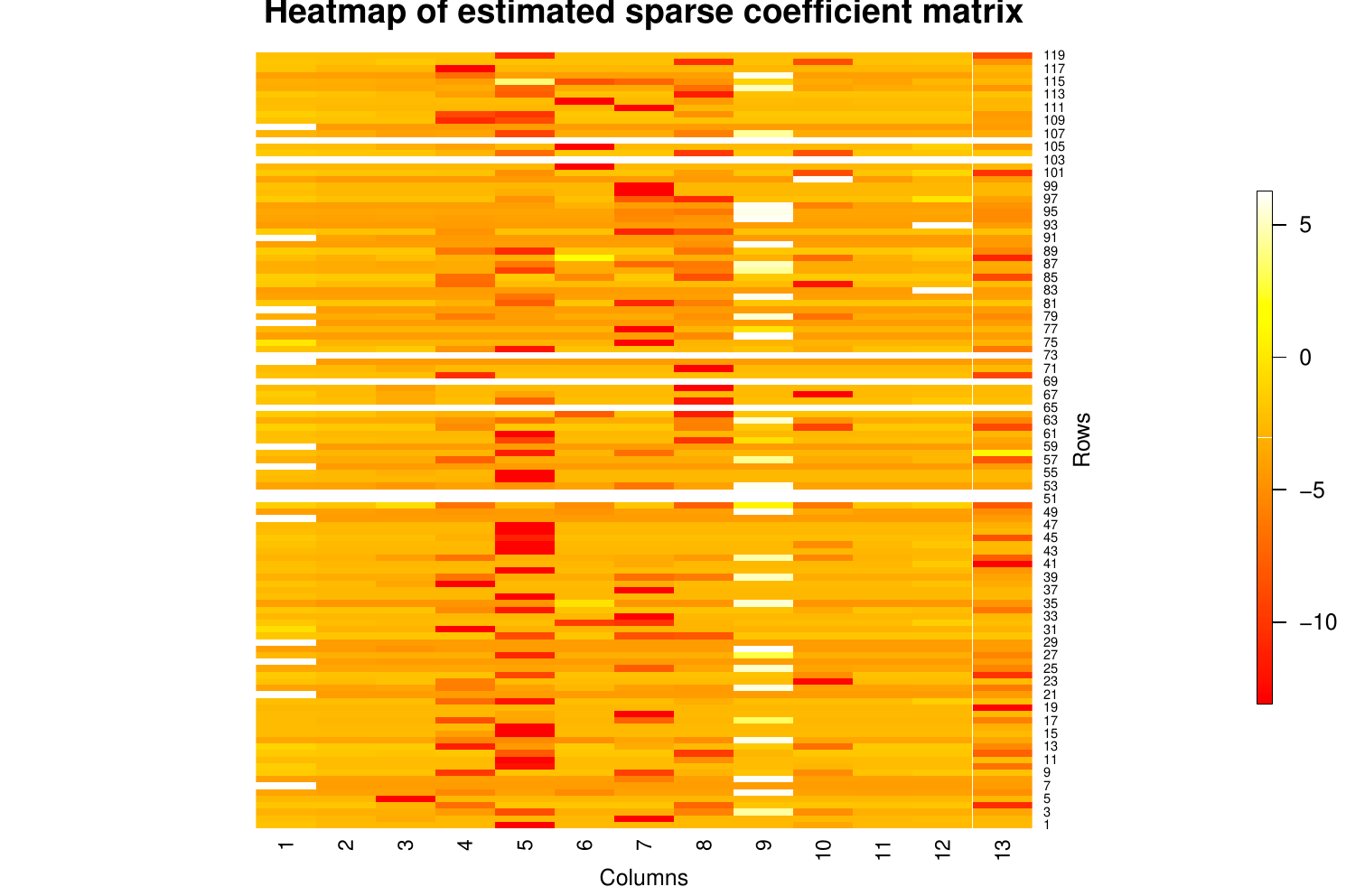}}
\caption{The heatmap of the estimated sparse coefficient matrix $\wh\bB$ in Example 3. }\label{fig8}
\end{center}
\end{figure}

 We first use Lasso to estimate the regression coefficient matrix. The heat map of the sparse matrix are shown in Figure~\ref{fig8}. From Figure~\ref{fig8}, we see that the largest average of the coefficients are produced by the risk-free interest rate. In other words, the dependence of stock returns 
 on the risk-free rate is the  strongest among the 13 predictors. Although the dependence on the dividends is weak for all stock returns from the heatmap, we find that the dividends predictors have more nonzero coefficients and there are 98 stocks which have dependence on the dividends. This is consistent with the findings in the finance literature that the dividends have significant predictive ability for the stock returns. See, for example, \cite{campbell1988}.

\begin{figure}
\begin{center}
{\includegraphics[width=0.8\textwidth]{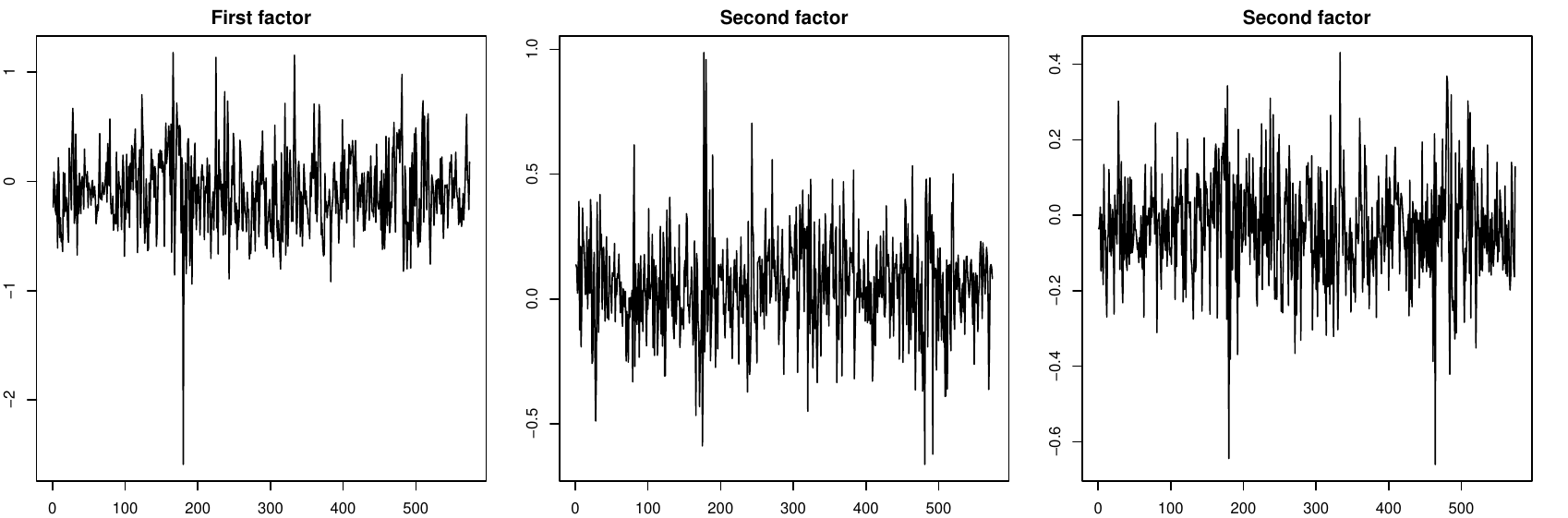}}
\caption{The time series plots of the three factors identified by the white noise test in Example 3. }\label{fig9}
\end{center}
\end{figure}

Next, we study whether there is a latent factor structure in the residuals. We apply the white noise test and find $\wh r =3$, i.e., there are 3 dynamically dependent factors embedded in the residuals. The time series plots of the three  factors are shown in Figure~\ref{fig9}. 
Although predictability of asset returns has been a controversial issue in the finance literature, we conduct a predictive analysis for the stock returns 
using the proposed model and procedure to explore the ability of 
predictive reressors and latent factors. We estimated the model in the time span $[1,\tau]$ with $\tau=576-T_0,576-T_0+1,...,575$ for the $1$-step ahead forecast, where $T_0$ is the forecast window in month. We study the overall performance by employing a VAR(1) model for the estimated factor process to compute the 1-step ahead predictions for $\bfeta_t$ and by combining it with the regression part to produce predictions of $\by_t$. 
The forecast error is defined as

\begin{equation}\label{fee}
 FE=\frac{1}{T_0}\sum_{\tau=576-T_0}^{575}\|\wh\by_{\tau+1}-\by_{\tau+1}\|_2/\sqrt{p},   
\end{equation}
where $p=119$. Table~\ref{Table4} summarizes the forecast errors. We can see that the proposed method produces slight better results than the one that only use the predictors (or regressors), indicating that the proposed method 
could be useful in predicting asset returns.
In general, it is not easy to produce accurate forecasts in asset returns and the improvements by the proposed method could have substantial implications to practitioners.

\begin{table}
 \caption{The 1-step ahead mean squared forecast errors for all 119 stock returns.}
          \label{Table4}
\begin{center}
 \setlength{\abovecaptionskip}{0pt}
\setlength{\belowcaptionskip}{3pt}

\begin{tabular}{c|cc}
\hline
$T_0$&Proposal&No Latent Factors\\
\hline
12&0.059&0.061\\
24&0.060&0.061\\
36&0.061&0.062\\
48&0.061&0.062\\
60&0.064&0.065\\
\hline
\end{tabular}
          \end{center}
\end{table}

\section{Concluding Remarks}\label{sec5}

This paper proposed a new approach to modeling high-dimensional time series data using penalized linear regression and latent factors. The model used marks 
an important extension of the traditional linear regression model with 
time series errors. 
The latent factors capture dynamic dependence of the data that is not explained by the regressors. In traditional multivariate regression contexts, the error term is typically assumed to be independent and identically distributed (i.i.d.) or a white noise process. However, this assumption is often violated in high-dimensional settings. To overcome the difficulty, we assumed the error term follows a latent factor structure, where the factors capture the dynamic information of the residuals. We also allowed the white noise effect to be strong, which corresponds to the case of low signal to noise ratio 
commonly seen in applications. The proposed approach not only extracts 
further dynamic information of the data but also can handle 
more complicated error structure of the data. 

\section*{Appendix: Proofs}
\renewcommand{\theequation}{A.\arabic{equation}}
\setcounter{equation}{0}
In this section, we use $C$ or $c$ as a generic constant the value of which may change at different
places.

{\bf Proof of Theorem 1.} When $m$ is finite,  by a similar argument as (A.2) in the supplement of \cite{gaotsay2018b},  we can show that
\begin{equation}\label{cov:z}
\left\|\frac{1}{T}\sum_{t=1}^T\bz_t\bz_t'-E(\bz_t\bz_t')\right\|_F=O_p(mT^{-1/2})=O_p(T^{-1/2}),
\end{equation}
and
\begin{equation}\label{z-eta}
\left\|\frac{1}{T}\sum_{t=1}^T\bz_t\bfeta_t'\right\|_F=O_p(m^{1/2}p^{1/2}T^{-1/2})=O_p(p^{1/2}T^{-1/2}).
\end{equation}

Note that
\begin{equation}\label{bhat}
(\wh\bB-\bB)'=\left(\frac{1}{T}\sum_{t=1}^T\bz_t\bz_t'\right)^{-1}\left(\frac{1}{T}\sum_{t=1}^T\bz_t\bfeta_t'\right).
\end{equation}
By Assumption \ref{a3}(i) and (\ref{cov:z})-(\ref{bhat}), we have
\begin{equation}\label{b:rate}
\|\wh\bB-\bB\|_F=O_p(p^{1/2}T^{-1/2}).
\end{equation}

When $m$ is large,  we derive the convergence rate for the Lasso estimator in (\ref{lasso:e}).  By the basic inequality in Lemma 6.1 of \cite{buhlmann2011}, 
\begin{equation}\label{basic}
\frac{1}{T}\sum_{t=1}^T(y_{i,t}-\wh\bb_i'\bz_t')^2+\lambda\|\wh\bb_i\|_1\leq \frac{1}{T}\sum_{t=1}^T(y_{i,t}-\bb_i'\bz_t)^2+\lambda\|\bb_i\|_1,
\end{equation}
which implies that
\begin{equation}\label{ineq}
\frac{1}{T}\sum_{t=1}^T[(\wh\bb_i-\bb_i)'\bz_t]^2\leq \frac{2}{T}\sum_{t=1}^T\eta_{i,t}\bz_t'(\wh\bb_i-\bb_i)+\lambda\{\|\bb_i\|_1-\|\wh\bb_i\|_1\}.
\end{equation}
Let $\wh\bDelta_i=\wh\bb_i-\bb_i$,  by the results in Lemma 6.3 of \cite{buhlmann2011}, we have $\wh\bDelta_i\in C_3(S_i)=\{\bDelta\in R^{m}:\|\bDelta_{S_i^c}\|_1\leq 3\|\bDelta_{S_i}\|_1\}$. By (\ref{ineq}), the restricted eigenvalue condition in Assumption \ref{a3}(ii),  H\"{o}lder's inequality, and the triangle inequality,
\begin{equation}\label{kappa}
\kappa\|\wh\bDelta_i\|_2^2\leq2\|\frac{1}{T}\sum_{t=1}^T\eta_{i,t}\bz_t'\|_{\infty}\|\wh\bDelta_i\|_1+\lambda\|\wh\bDelta_i\|_1.
\end{equation}

By Assumptions \ref{a1}--\ref{a2}, Lemma 3 in \cite{fan2013}, and Theorem 1 in \cite{merlevede2011},  there exists a $\gamma>0$ such that
\begin{align}\label{max:in}
P(\max_{1\leq i\leq p, 1\leq j\leq m}\left|\frac{1}{T}\sum_{t=1}^T\eta_{i,t}z_{j,t}\right|>x)\leq& \sum_{i=1}^p\sum_{j=1}^mP(\left|\frac{1}{T}\sum_{t=1}^T\eta_{i,t}z_{j,t}\right|>x)\notag\\
\leq &pmT\exp(\frac{(Tx)^\gamma}{C})+pm\exp(-\frac{(Tx)^2}{CT})\notag\\
&+pm\exp\left(-\frac{(Tx)^2}{CT}\exp(\frac{(Tx)^{\gamma(1-\gamma)}}{C(\log(Tx))^\gamma})\right),
\end{align}
which implies that 
\[\max_{1\leq i\leq p, 1\leq j\leq m}\left|\frac{1}{T}\sum_{t=1}^T\eta_{i,t}z_{j,t}\right|=O_p(\sqrt{\frac{\log(pm)}{T}}).\]
Therefore, we have $\max_{1\leq i\leq p, 1\leq j\leq m}\left|\frac{1}{T}\sum_{t=1}^T\eta_{i,t}z_{j,t}\right|\le M\sqrt{\frac{\log(pm)}{T}}$ for a sufficiently large $M>0$ with probability tending to one. On the event $\{\max_{1\leq i\leq p, 1\leq j\leq m}\left|\frac{1}{T}\sum_{t=1}^T\eta_{i,t}z_{j,t}\right|\le M\sqrt{\frac{\log(pm)}{T}}\}$, for any $\lambda_T\geq M\sqrt{\frac{\log(pm)}{T}}$,  it follows from (\ref{kappa})  that
\begin{align}\label{delta:in}
\kappa\|\wh\bDelta_i\|_2^2\leq &3\lambda_T\|\wh\bDelta_i\|_1\leq 3\lambda_T(\|\wh\bDelta_{S_i}\|_1+\|\wh\bDelta_{S_i^c}\|_1)\notag\\
\leq & 3\lambda_T(\|\wh\bDelta_{S_i}\|_1+3\|\wh\bDelta_{S_i}\|_1)
\leq  12\lambda_T\|\wh\bDelta_{S_i}\|_1\leq 12\sqrt{s_i}\lambda_T\|\wh\bDelta_i\|_2,
\end{align}
which implies that
\[\|\wh\bDelta_i\|_2\leq 12\frac{\sqrt{s^*}}{\kappa}\lambda_T,\]
uniformly for $1\leq i\leq p$, where $s^*=\max\{s_i,i=1,...,m\}$.
Therefore,
\begin{equation}\label{B:lasso}
\|\wh\bB-\bB\|_F\leq 12\frac{\sqrt{ps^*}}{\kappa}\lambda_T.
\end{equation}
This completes the proof of Theorem 1. $\Box$

\begin{lemma}\label{lm1}
If Assumptions \ref{a1}, \ref{a2}, \ref{a4} and \ref{a5} hold, then
\[\|\bSigma_\eta(k)\|_2=O_p(p^{1-\delta_1}+\kappa_{\max}p^{1-\delta_1/2-\delta_2/2})\quad\text{for}\quad 1\leq k\leq k_0,\]
and
\[\|\bSigma_\eta\|_2=O_p(p^{1-\delta_1}+p^{1-\delta_2}).\]
\end{lemma}

{\bf Proof. } It is the same as Lemma 3 in \cite{gaotsay2018b}. We omit the details. $\Box$

\begin{lemma}\label{lm2}
 If Assumptions \ref{a1}--\ref{a5} hold, then, for $0\leq k\leq k_0$, if $p^{\delta_1/2}T^{1/2}=o(1)$ and $p^{\delta_2/2}T^{1/2}=o(1)$ when $m$ is finite, and $p^{\delta_1/2}\log(pm)T^{-1/2}=o(1)$ and $p^{\delta_1/2}\log(pm)T^{-1/2}=o(1)$ when $m$ is diverging, we have
\[ \|\wh\bSigma_\eta(k)-\bSigma_\eta(k)\|_2=\left\{\begin{array}{ll}
O_p(p^{1-\delta_1/2}T^{-1/2}),&\text{if}\,\, \delta_1\leq \delta_2,\\
O_p(p^{1-\delta_2/2}T^{-1/2}),&\text{if}\,\, \delta_1> \delta_2.
\end{array}\right.\]
\end{lemma}

{\bf Proof.} By definition, 
\begin{align}\label{sig:diff}
\wh\bSigma_\eta(k)-\bSigma_\eta(k)=&\frac{1}{T}\sum_{t=k+1}^T(\wh\bfeta_t-\bar{\bfeta})(\wh\bfeta_{t-k}-\bar{\bfeta})'-E(\bfeta_t\bfeta_{t-k}')\notag\\
=&\frac{1}{T}\sum_{t=k+1}^T\{\wh\bfeta_t\wh\bfeta_{t-k}'-\bfeta_t\bfeta_{t-k}'\}+\left\{\frac{1}{T}\sum_{t=k+1}^T\bfeta_t\bfeta_{t-k}'-E(\bfeta_t\bfeta_{t-k}')\right\}\notag\\
&+\bar{\bfeta}\bar{\bfeta}'-\frac{1}{T}\sum_{t=k+1}^T\wh\bfeta_t\bar{\bfeta}'-\frac{1}{T}\sum_{t=k+1}^T\bar{\bfeta}\wh\bfeta_{t-k}'\notag\\
=&I_{1,k}+I_{2,k}+I_{3,k}+I_{4,k}+I_{5,k}.
\end{align}

 We first consider the case when $m$ is finite.  Note that $\wh\bfeta_t=\by_t-\wh\bB\bz_t=\bfeta_t+(\bB-\wh\bB)\bz_t$, it follows that
 \begin{align}\label{i1}
 I_{1,k}=&\frac{1}{T}\sum_{t=k+1}^T(\bB-\wh\bB)\bz_t\bz_{t-k}'(\bB-\wh\bB)'+\frac{1}{T}\sum_{t=1}^T(\bB-\wh\bB)\bz_t\bfeta_{t-k}'\notag\\
 &+\frac{1}{T}\sum_{t=k+1}^T\bfeta_t\bz_{t-1}'(\bB-\wh\bB)'.
 \end{align}
Therefore,
\begin{align}\label{i1:norm}
\|I_{1,k}\|_2\leq &\|\bB-\wh\bB\|_F^2\|\frac{1}{T}\sum_{t=k+1}^T\bz_t\bz_{t-k}'\|_2++\|\bB-\wh\bB\|_F\|\frac{1}{T}\sum_{t=k+1}^T\frac{1}{T}\bz_t\bfeta_{t-k}\|_2\notag\\
&+\|\bB-\wh\bB\|_F\|\frac{1}{T}\sum_{t=k+1}^T\bfeta_t\bz_{t-k}\|_2\notag\\
\leq &O_p(\|\bB-\wh\bB\|_F^2)O_p(1)+O_p(\|\bB-\wh\bB\|_F)O_p(\sqrt{\frac{p}{T}})+O_p(\|\bB-\wh\bB\|_F)O_p(\sqrt{\frac{p}{T}})\notag\\
\leq &O_p(pT^{-1}).
\end{align}
By Assumptions \ref{a1}--\ref{a5} and a similar argument as Lemma 4(ii) in \cite{gaotsay2018b}, we can show that
\begin{equation}\label{i2}
\\|I_{2,k}\|_2=\left\{\begin{array}{ll}
O_p(p^{1-\delta_1/2}T^{-1/2}),&\text{if}\,\, \delta_1\leq \delta_2,\\
O_p(p^{1-\delta_2/2}T^{-1/2}),&\text{if}\,\, \delta_1> \delta_2.
\end{array}\right.
\end{equation}
Similarly, we can show that
\begin{equation}\label{i345}
\|I_{3,k}\|_2=O_p(pT^{-1}), \|I_{4,k}\|_2=O_p(pT^{-1}),\,\,\text{and}\,\,\|I_{5,k}\|_2=O_p(pT^{-1}).
\end{equation}
Thus, if $p^{\delta_1/2}T^{1/2}=o(1)$ and $p^{\delta_2/2}T^{1/2}=o(1)$, Lemma 2 follows from (\ref{i1})-(\ref{i345}).

Now, we consider the case when $m$ is diverging.  By Schwarz's inequality and Theorem 6.1 in \cite{buhlmann2011}, 

\begin{equation}\label{bijz}
|\frac{1}{T}\sum_{t=k+1}^T(\bb_i-\wh\bb_i)'\bz_t\bz_{t-k}'(\bb_j-\wh\bb_j)'|\leq C\frac{1}{T}\sum_{t=k+1}^T[(\bb_i-\wh\bb_i)'\bz_t]^2\leq C\lambda_T^2s^*,
\end{equation}
uniformly for $1\leq i,j\leq p$. Therefore,
\begin{equation}\label{i1:1}
\|\frac{1}{T}\sum_{t=k+1}^T(\bB-\wh\bB)'\bz_t\bz_{t-k}'(\bB-\wh\bB)'\|_2=O_p(\sqrt{p^2\lambda_T^4s^{*2}})=O_P(p\lambda_T^2s^*)=O_(\frac{ps^*\log(pm)}{T}).
\end{equation}
Furthermore, by H\"{o}lder's inequality and (\ref{B:lasso}) or Theorem 6.1 of \cite{buhlmann2011},  
\begin{align}\label{i1:2}
\|\frac{1}{T}\sum_{t=k+1}^T(\bB-\wh\bB)'\bz_t\bfeta_{t-k}\|_2\leq& \|\bB-\wh\bB\|_\infty\|\frac{1}{T}\sum_{t=k+1}^T\bz_t\bfeta_{t-k}'\|_1\notag\\
\leq &O_p(\lambda_Ts^*)O_p(mT^{-1/2})=O_p(\frac{ms^*\sqrt{\log(pm)}}{T}),
\end{align}
and similar result holds for the third term of $I_{1,k}$ in (\ref{i1}).  As $m\leq Cp$, we have
\[\|I_{1,k}\|_2=O_p(\frac{ps^*\log(pm)}{T}).\]
As for $I_{2,k}$, the result in (\ref{i2}) still holds. By a similar argument as (\ref{bijz})-(\ref{i1:2}),  if $m\leq Cp$ and $m\leq T$, we can show that
\begin{equation}\label{i345:a}
\|I_{3,k}\|_2=O_p(\frac{ps^*\log(pm)}{T}), \|I_{4,k}\|_2=O_p(\frac{ps^*\log(pm)}{T}), \|I_{5,k}\|_2=O_p(\frac{ps^*\log(pm)}{T}).
\end{equation}
If $p^{\delta_1/2}s^*\log(pm)T^{-1/2}=o(1)$ and $p^{\delta_1/2}s^*\log(pm)T^{-1/2}=o(1)$, we still obtain the results in Lemma \ref{lm2}. This completes the proof. $\Box$

Lemma \ref{lm3}--\ref{lm6} below are similar to Lemmas 5--7 in the Supplement of \cite{gaotsay2018b} and we only briefly state them without giving the detailed proofs.
\begin{lemma}\label{lm3}
If Assume Assumptions \ref{a1}--\ref{a5} hold,  $p^{\delta_1/2}T^{1/2}=o(1)$ and $p^{\delta_2/2}T^{1/2}=o(1)$ when $m$ is finite, and $p^{\delta_1/2}\log(pm)s^*T^{-1/2}=o(1)$ and $p^{\delta_1/2}s^*\log(pm)T^{-1/2}=o(1)$ when $m$ is diverging.  If $\delta_1 \leq \delta_2$, then
\begin{equation*}
\|\wh\bM-\bM\|_2=O_p(p^{2-3\delta_1/2}n^{-1/2}+\kappa_{\max}p^{2-\delta_1-\delta_2/2}n^{-1/2}).
\end{equation*}
If $\delta_1>\delta_2$, then
\begin{equation*}
\|\wh\bM-\bM\|_2=\left\{\begin{array}{ll}
O_p(p^{2-\delta_2}n^{-1}+p^{2-\delta_1-\delta_2/2}n^{-1/2}),& \text{if}\,\,\kappa_{\max}=0,\\
O_p(\kappa_{\max} p^{2-\delta_1/2-\delta_2}n^{-1/2}),&\text{if} \,\, \kappa_{\max}>>0.
\end{array}\right.
\end{equation*}
\end{lemma}
\begin{lemma}\label{lm4}
If Assumptions \ref{a1}--\ref{a5} hold, then
\begin{equation}\label{lbdr}
\lambda_{\min}(\bM)\geq\left\{\begin{array}{ll}
Cp^{2(1-\delta_1)},&\text{if}\,\, \kappa_{\max}p^{\delta_1/2-\delta_2/2}=o(1),\\
C\kappa_{\min}^2p^{2-\delta_1-\delta_2},& \text{if}\,\, r\leq s\,\,\text{and}\,\, \kappa_{\min}^{-1}p^{\delta_2/2-\delta_1/2}=o(1),\\
C\kappa_{\min}^2 p^{1-\delta_1},&\text{if}\,\, r>s\,\,\text{and}\,\, \kappa_{\min}^{-1}p^{(1-\delta_1)/2}=o(1),
\end{array}\right.
\end{equation}
where $\kappa_{\min}$ and $\kappa_{\max}$ are defined in (\ref{kmm}) and $s$ is given in Assumption \ref{a5}.
\end{lemma}

\begin{lemma}\label{lm5}
If Assumptions \ref{a1}--\ref{a5} hold, then
\[\lambda_{s}(\bS)\geq Cp^{2-2\delta_2}.\]
\end{lemma}

\begin{lemma}\label{lm6}
 Let Assumptions \ref{a1}--\ref{a5} hold, $p^{\delta_1/2}T^{1/2}=o(1)$ and $p^{\delta_2/2}T^{1/2}=o(1)$ when $m$ is finite, and $p^{\delta_1/2}\log(pm)s^*T^{-1/2}=o(1)$ and $p^{\delta_1/2}s^*\log(pm)T^{-1/2}=o(1)$ when $m$ is diverging.   If $\delta_1\leq \delta_2$, then
\[\|\wh\bS-\bS\|_2\leq Cp^{2-3\delta_1/2}n^{-1/2}+Cp^{2-2\delta_1}\|\wh\bU_1-\bU_1\|_2.\]
If $\delta_1>\delta_2$,
\[\|\wh\bS-\bS\|_2\leq Cp^{2-3\delta_2/2}n^{-1/2}+Cp^{2-2\delta_2}\|\wh\bU_1-\bU_1\|_2.\]
\end{lemma}

{\bf Proof of Theorem 2.} Letting $\bA=\bM$ and $\bE=\wh\bM-\bM$ in Lemma 1, we can obtain
\[\|\wh\bA_1-\bA_1\|_2\leq \frac{\|\wh\bM-\bM\|_2}{\lambda_{\min}(\bM)},\,\,\|\wh\bU_1-\bU_1\|_2\leq \frac{\|\wh\bM-\bM\|_2}{\lambda_{\min}(\bM)}\]
and
\[\|\wh{\bU}_2^*-\bU_2^*\|_2\leq \frac{\|\wh\bS-\bS\|_2}{\lambda_{s}(\bS)}.\]
Theorem 2 can then be shown by an elementary argument based on Lemmas 3-6. 
We omit the details. This completes the proof. $\Box$

{\bf Proof of Theorem 3.} Note that
\begin{align}\label{ax:dif}
\wh\bA_1\wh\bx_t-\bA_1\bx_t=
&[\wh\bA_1(\wh\bU_2'\wh\bA_1)^{-1}\wh\bU_2'\bfeta_t-\bA_1\bx_t]+\wh\bA_1(\wh\bU_2'\wh\bA_1)^{-1}\wh\bU_2'(\wh\bfeta_t-\bfeta_t)\\
=&J_1+J_2.\notag
\end{align}
By a similar argument as the proof of Theorem 4 in \cite{gaotsay2018b}, we have
\[p^{-1/2}\|J_1\|_2=O_p(p^{-1/2}+p^{-\delta_1/2}D(\wh\bA_1,\bA_1)+p^{-\delta_2/2}D(\wh\bU_2^*,\bU_2^*)).\]
Since $\bfeta_t-\wh\bfeta_t=(\wh\bB-\wh\bB)\bz_t$,  by Theorem 1, 
\[p^{-1/2}\|J_2\|_2=p^{-1/2}\|\bB-\wh\bB\|_F\|\bz_t\|_2=O_p(\sqrt{\frac{ms^*\log(pm)}{T}}).\]
Thus,
\[p^{-1/2}\|\wh\bA_1\wh\bx_t-\bA_1\bx_t\|_2=O_p(p^{-1/2}+p^{-\delta_1/2}D(\wh\bA_1,\bA_1)+p^{-\delta_2/2}D(\wh\bU_2^*,\bU_2^*)+\sqrt{\frac{ms^*\log(pm)}{T}}).\]
This completes the proof.  $\Box$

{\bf Proof of Theorem 4.} Note that
\begin{equation}\label{uy}
\wh\bU_1'\wh\bfeta_t=(\wh\bU_1-\bU_1)'\bfeta_t+\wh\bU_1'(\bB-\wh\bB)\bz_t+\bU_1'\bA_2\bve_t.
\end{equation}
As each component of $\bU_1'\bA_2\bve_t$ is $O_p(1)$, we only need to show that each component of the first and the second term in (\ref{uy}) is of $o_p(1)$ such that the vector in (\ref{uy}) is asymptotically a white noise. That is, we require
\[\max_{1\leq  i\leq p-\wh r,1\leq t\leq T}|(\wh\bu_{1,i}-\bu_{1,i})'\bfeta_t|=o_p(1)\,\,\text{and}\,\,\max_{1\leq  i\leq p-\wh r,1\leq t\leq T}|\wh\bu_{1,i}'(\bB-\wh\bB)\bz_t|=o_p(1).\]
By Assumption 2, 
\[P(\max_{1\leq t\leq T}\|\bfeta_t\|_2\geq x)\leq TpP(\eta_{i,t}|\geq x/\sqrt{p})\leq CTp\exp(-C(x/\sqrt{p})^{\gamma_2}),\]
which implies that $\max_{1\leq t\leq T}\|\bfeta_t\|_2=O_p(\sqrt{p}(\log(Tp))^{1/\gamma_2})$. Similarly, we can show that $\max_{1\leq t\leq T}\|\bz_t\|_2=O_p(\sqrt{m}(\log(Tm))^{1/\gamma_2})$. 
By Assumption 2 and the proof of Theorem 1, we have
\[\max_{1\leq  i\leq p-\wh r,1\leq t\leq T}|(\wh\bu_{1,i}-\bu_{1,i})'\bfeta_t|\leq CD(\wh\bU_1,\bU_1)\max_{1\leq t\leq T}\|\bfeta_t\|_2=O_p(\sqrt{p}(\log(Tp))^{1/\gamma_2}D(\wh\bU_1,\bU_1)),\]
and
\[\max_{1\leq  i\leq p-\wh r,1\leq t\leq T}|\wh\bu_{1,i}'(\bB-\wh\bB)\bz_t|\leq C\|\bB-\wh\bB\|_2\max_{1\leq t\leq T}\|\bz_t\|_2=O_p(\sqrt{\frac{pms^*\log(pm)}{T}}(\log(Tm))^{1/\gamma_2}). \]
We only need to require the above rate to be $o(1)$. This completes the proof. $\Box$


\begin{thebibliography}{99}



\bibitem[{Ahn and Horenstein(2013)}]{ahn2013}
Ahn, S. C., and Horenstein, A. R. (2013). Eigenvalue ratio test for the number of factors.
{\em Econometrica}, {\bf 81(3)}, 1203–d1227.

\bibitem[{Anderson(2003)}]{anderson2003}
Anderson, T. W. (2003). {\it An Introduction to Multivariate Statistical Analysis,}
Hoboken,  NJ: Wiley.






\bibitem[{Bai(2003)}]{Bai_Econometrica_2003}
Bai J. (2003) Inferential theory for factor models of large dimensions. \textsl{Econometrica}, {\bf 71(1)}, 135--171.



\bibitem[{Bai and Ng(2002)}]{BaiNg_Econometrica_2002}
Bai, J. and Ng, S. (2002). Determining the number of factors in approximate factor models. \textsl{Econometrica}, {\bf 70}, 191--221.



\bibitem[{Black(1986)}]{black1986}
Black, F. (1986). 
\newblock Noise. \newblock {\em The Journal of Finance}, {\bf 41(3)}, 528--543.



\bibitem[{Box and Tiao(1977)}]{BoxTiao_1977}
Box, G. E. P. and Tiao, G. C. (1977). A canonical analysis of multiple time series. {\sl Biometrika}, {\bf 64}, 355--365.



\bibitem[{B\"{u}hlmann and Van De Geer(2011)}]{buhlmann2011}
B\"{u}hlmann, P., and Van De Geer, S. (2011).  {\it Statistics for High-dimensional Data: Methods,  Theory and Applications.  }
Springer Science \& Business Media.


\bibitem[Chen~et~al., 2020]{chentsaychen2018}
Chen, E.Y., Tsay, R.S., and Chen, R. (2020). 
\newblock Constrained factor models for high-dimensional matrix-variate time series. 
\newblock {\em Journal of the American Statistical Association}, {\bf 115(530)}, 775--793.



\bibitem[{Davis{ et al.}(2016)}]{Davis2012}
Davis, R. A., Zang, P., and Zheng, T. (2016). Sparse vector autoregressive modeling. {\em Journal of Computational and Graphical Statistics}, {\bf 25(4)}, 1077--1096.

\bibitem[Campbell and Shiller, 1988]{campbell1988}
Campbell, J. Y., and Shiller, R. J. (1988). The dividend-price ratio and expectations of future dividends and discount factors. {\it The Review of Financial Studies}, 1(3), 195--228.

\bibitem[Chang~et~al., 2015]{changguoyao2015}
Chang, J., Guo, B. and Yao, Q. (2015). High dimensional stochastic regression with latent factors, endogeneity and nonlinearity. {\it Journal of Econometrics}, {\bf 189(2)}, 297--312.



\bibitem[{Fama and French(2015)}]{famafrench2015}
Fama, E. F. and French, K. R. (2015). A five-factor asset pricing model. 
\newblock {\em Journal of Financial Economics}, {\bf 116(1)}, 1--22.



\bibitem[{Fan {et al.}(2013)}]{fan2013}
Fan, J., Liao, Y., and Mincheva, M. (2013). Large covariance estimation by thresholding principal orthogonal complements (with discussion). 
\newblock {\em Journal of the Royal Statistical Society}, Series B, {\bf 75(4)}, 603--680.


\bibitem[{Forni { et al.}(2000)}]{forni2000}
Forni, M., Hallin, M., Lippi, M. and Reichlin, L. (2000).
\newblock Reference cycles: the NBER methodology revisited (No. 2400). Centre for Economic Policy Research.

\bibitem[{Forni { et al.}(2005)}]{forni2005}
Forni, M., Hallin, M., Lippi, M. and Reichlin, L. (2005). 
\newblock The generalized dynamic factor model: one-sided estimation and forecasting. 
\newblock {\em Journal of the American Statistical Association}, {\bf 100(471)}, 830--840.


\bibitem[{Gao{ et al.}(2019)}]{gaoetal2017}
Gao, Z., Ma, Y., Wang, H. and Yao, Q. (2019). 
\newblock Banded spatio-temporal autoregressions. 
\newblock {\em Journal of Econometrics,} {\bf 208(1)}, 211--230.


\bibitem[{Gao and Tsay(2019)}]{gaotsay2018a}
Gao, Z. and Tsay, R. S. (2019). 
\newblock A structural-factor approach for modeling high-dimensional time series and space-time data. 
\newblock {\em Journal of Time Series Analysis}, {\bf 40}, 343--362.

\bibitem[{Gao and Tsay(2021)}]{gaotsay2020a}
Gao, Z. and Tsay, R. S. (2021). 
\newblock Modeling high-dimensional unit-root time series.
\newblock {\em International Journal of Forecasting}, {\bf 37(4)},  1535--1555.


\bibitem[{Gao and Tsay(2022)}]{gaotsay2018b}
Gao, Z. and Tsay, R. S. (2022). 
\newblock Modeling high-dimensional time series: a factor model with
dynamically dependent factors and diverging eigenvalues.
\newblock  {\em Journal of the American Statistical Association}, {\bf 117(539)}, 1398--1414.

\bibitem[{Gao and Tsay(2023a)}]{gaotsay2021c}
Gao, Z. and Tsay, R. S. (2023a). 
\newblock A two-way transformed factor model for
matrix-variate time series.
\newblock  {\em Econometrics and Statistics}, {\bf 27}, 83--101.

\bibitem[{Gao and Tsay(2023b)}]{gaotsay2021}
Gao, Z. and Tsay, R. S. (2023b). 
\newblock Divide-and-conquer: a distributed hierarchical factor approach to modeling large-scale time series data.
\newblock  {\em Journal of the American Statistical Association}, {\bf 118(544)}, 2698--2711.










\bibitem[{Han {et al.}(2020)}]{Hanetal2020}
Han, Y., Chen, R., Yang, D., and Zhang, C. H. (2020). Tensor factor model estimation by iterative projection. {\em arXiv preprint arXiv:2006.02611}.











\bibitem[Lam and Yao(2012)]{lamyao2012}
Lam, C. and Yao, Q. (2012).
\newblock Factor modeling for high-dimensional time series: inference for the number of factors.
\newblock {\em The Annals of Statistics}, {\bf 40(2)}, 694--726.




\bibitem[{Lam {et al.}(2011)}]{LamYaoBathia_Biometrika_2011}
Lam, C., Yao, Q. and Bathia, N. (2011). Estimation of latent factors for high-dimensional time series. \BKA, {\bf98}, 901--918.






\bibitem[{Merlev\`{e}de {et al.}(2011)}]{merlevede2011}
Merlev\`{e}de, F., Peligrad, M. and Rio, E. (2011). A Bernstein type inequality and moderate
deviations for weakly dependent sequences. {\em Probability Theory and Related Fields}, {\bf 151(3)}, 435-–474.



\bibitem[Onatski, 2010]{onatski2010}
Onatski, A. (2010). \newblock Determining the number of factors from empirical distribution of eigenvalues. 
\newblock {\em The Review of Economics and Statistics}, {\bf 92(4)}, 1004--1016.

\bibitem[Pan and Yao, 2008]{panyao2008}
Pan, J. and Yao, Q. (2008).
\newblock Modelling multiple time series via common factors.
\newblock {\em Biometrika}, {\bf 95(2)}, 365--379.





\bibitem[Sharpe, 1964]{sharpe1964}
Sharpe, W. F. (1964). Capital asset prices: A theory of market equilibrium under conditions of risk. 
\newblock {\em The Journal of Finance}, {\bf 19(3)}, 425--442.



\bibitem[{Shojaie and Michailidis(2010)}]{ShojaieMichailidis_2010}
Shojaie, A. and Michailidis, G. (2010). Discovering graphical Granger causality using the truncated lasso penalty. \textsl{Bioinformatics}, {\bf 26}, 517--523.



\bibitem[{Song and Bickel(2011)}]{SongBickel_2011}
Song, S. and Bickel, P. J. (2011). Large vector auto regressions. Available at {\sl arXiv:1106.3519}.




\bibitem[{Stewart and Sun(1990)}]{stewart-sun1990}
Stewart, G. W., and Sun, J. (1990). {\it Matrix Perturbation Theory}. Academic Press.


\bibitem[{Stock and Watson(2002)}]{StockWatson_2002}
Stock, J. H. and Watson, M. W. (2002). Forecasting using principal components from a large number of predictors. \JASA, {\bf 97}, 1167--1179.


\bibitem[{Stock and Watson(2005)}]{StockWatson_2005}
Stock, J. H. and Watson, M. W. (2005). Implications of dynamic factor models for VAR analysis. NBER Working Paper 11467.



\bibitem[{Tiao and Tsay(1989)}]{TiaoTsay_1989}
Tiao, G. C. and Tsay, R. S. (1989). Model specification in multivariate time series (with discussion). 
\newblock {\em \JRSSB}{\bf51}, 157--213.


 \bibitem[{Tsay(2014)}]{Tsay_2014}
 Tsay, R. S. (2014). {\sl Multivariate Time Series Analysis}. Wiley, Hoboken, NJ.

 \bibitem[{Tsay(2020)}]{Tsay_2018}
Tsay, R. S. (2020). 
\newblock Testing for serial correlations in high-dimensional time series via extreme value
theory. \newblock {\em Journal of Econometrics}, {\bf 216}, 106–-117.



 \bibitem[{Wainwright(2019)}]{Wainwright2019}
Wainwright, M. J. (2019). {\sl High-dimensional Statistics: A Non-asymptotic Viewpoint.} Cambridge University Press.

 \bibitem[{Wang~et~al.(2019)}]{wang2018}
Wang, D., Liu, X. and Chen, R. (2019). 
\newblock Factor models for matrix-valued high-dimensional time series. 
\newblock {\em Journal of Econometrics}, {\bf 208(1)}, 231--248.

\bibitem[{Wang~et~al.(2021)}]{wangzhengli2021}
Wang, D., Zheng, Y., and Li, G. (2021).
\newblock High-dimensional low-rank tensor autoregressive time series modeling. 
{\em Journal of Econometrics}, {\bf 238(1)}, 105544.

\bibitem[{Wang~et~al.(2020)}]{wanglianzhengli2020}
Wang, D., Zheng, Y., Lian, H., and Li, G. (2022).
\newblock High-dimensional vector autoregressive time series modeling via tensor decomposition.
\newblock{\em Journal of the American Statistical Association}, {\bf 117(539)}, 1338-1356.


 \bibitem[Welch and Goyal, 2008]{welch2008}
Welch, I., and Goyal, A. (2008). A comprehensive look at the empirical performance of equity premium prediction. {\it The Review of Financial Studies}, {\bf 21(4)}, 1455--1508.









\end{thebibliography}
\end{document}